\documentclass[preprint, 12pt,3p]{elsarticle}
\usepackage{lineno,hyperref}
\modulolinenumbers[5]

\usepackage{gensymb}
\usepackage{eurosym}
\usepackage{color}
\usepackage{amsmath}
\usepackage{graphicx}
\usepackage{nomencl}
\usepackage{multicol}

\makenomenclature

\usepackage{etoolbox}
\renewcommand\nomgroup[1]{%
  \item[\bfseries
  \ifstrequal{#1}{C}{Power Conversion}{%
  \ifstrequal{#1}{R}{Resource Analysis}{%
  \ifstrequal{#1}{O}{Network Optimisation}{}}}%
]}

\journal{Applied Energy}

\begin{document}
\begin{frontmatter}
\title{The impact of climate change on a cost-optimal highly renewable European electricity network}
\author[fias]{Markus~Schlott}

\author[fias]{Alexander~Kies}

\author[fias,kit]{Tom~Brown}

\author[fias]{Stefan~Schramm}

\author[aarhus]{Martin~Greiner}
\address[fias]{Frankfurt Institute for Advanced Studies, Goethe University Frankfurt, Ruth-Moufang-Str. 1, 60438 Frankfurt am Main, Germany}
\address[aarhus]{Department of Engineering,  Aarhus University, Inge Lehmans Gade 10,
8000 Aarhus C, Denmark}
\address[kit]{Institute for Automation and Applied Informatics, Karlsruhe Institute of Technology, 76344 Eggenstein-Leopoldshafen, Germany}

\begin{abstract}

We use three ensemble members of the EURO-CORDEX project and their data on surface wind speeds, solar irradiation as well as water runoff with a spatial resolution of 12 km and a temporal resolution of 3 hours under representative concentration pathway 8.5 (associated with a strong climate change and a temperature increase of 2.6 to 4.8 \degree C until the end of the century) until 2100 to investigate the impact of climate change on wind, solar and hydro resources and consequently on a highly renewable and cost-optimal European power system. \\
The weather data is transformed into power, different aspects such as capacity factors and correlation lengths are investigated and the resulting implications for the European power system are discussed. In addition, we compare a 30-node model of Europe with historical and climate change-affected data, where investments in generation, transmission and storage facilities are optimised. \\
Differences in capacity factors among European countries are more strongly emphasized at the end of the century compared to historic data.
This results in a significantly increased photovoltaic share in the cost-optimal power system. In addition, annual hydro inflow patterns of major hydro producers change considerably.
System costs increase by 5\% until the end of the century and the impact of climate change on these costs is of similar magnitude as differences between the ensemble members. 
The results show that including climate affected-weather data in power system simulations of the future has an observeable effect.
\end{abstract}

\begin{keyword}
Renewable Power System, Climate Change, EURO-CORDEX, Wind Resources, Solar Resources, Hydro Resources, Energy System Analysis
\end{keyword}
\end{frontmatter}


\section{Introduction}
Global warming is the century-scale rise in the average temperature of the Earth's climate system that has been observed and attributed to anthropogenic emissions of greenhouse gases such as carbon dioxide (CO$_2$). The associated statistical relevant changes in the weather patterns over long periods of time are referred to as climate change.
A global climate change induced by greenhouse gas emissions may have profound impact on the availability of renewable energy sources and consequently on the optimal design of future power systems. On the other hand, the large-scale deployment of renewable generation facilities in such power systems may also have the potential of keeping the global temperature increase below the officially set 2 \degree C goal \cite{rogelj2016paris}.
Besides, V{\'a}zquez-Rowe et al. \cite{vazquez2015climate} conclude that climate-centric policy-making for national electricity mixes can lead to substantial environmental co-benefits. Among other influences of climate change is possibly also water shortage, which directly influences thermal generation plants as pointed out by Zheng et al. \cite{zheng2016vulnerability}. 

In a recent review paper on trends and gaps in research on climate change science by 
Cronin et al. \cite{cronin2018climate}, the authors have derived the recommendation to investigate technical and cost impacts of altered variability of
renewable resources, and the impacts of extreme weather events on all elements of the energy
system. 
Moemken et al. \cite{moemken2018future} investigate future changes in wind energy potentials using EURO-Cordex data, find them to be robust over different ensemble members for intraannual variability and more uncertain for interdaily variability and emphasize the importance of research on the effect of climate change on the energy system at the European scale.
This study directly contributes to this aspect of climate change research. 

To study the effect of climate change on renewable power systemes,
a careful investigation of the requirement for all system elements is necessary, because the integration of renewable energy sources into power systems is, due to their weather-dependent variability, a challenging task \cite{lund2007renewable}.
Research often proposes the combination of different renewables to account for intermittency of single sources: 
For instance, Santos-Alamillos et al. \cite{santos2015combining} propose to combine wind and solar power for stable power output and Jurasz et al. \cite{jurasz2017integrating} to smoothen PV production by the use of hydro power.

Several studies have studied different influences of climate change on renewable power using different methodologies: 
Tobin et al. \cite{tobin2016climate} use EURO-CORDEX data and find that southern European wind power locations are strongly affected, but for most European regions differences in wind power potential are within $\pm 5\%$.
Ven{\"a}l{\"a}inen et al. \cite{venalainen2004influence} find wind power potential in Finland to increase by 2\%-10\% and
Pryor et al. \cite{pryor2005climate} find, using data from five downscaled general circulation models, decreasing mean wind speeds in Northern Europe until 2100 and increases in extreme wind speeds \cite{pryor2010climate}.
The effect of global warming on offshore winds on the Western Iberian Peninsula is studied by Soares et al. \cite{soares2017western} concluding decreases up to 20\% in winter and autumn and increases up to 20\% in summer.
Fant et al. \cite{fant2016impact} find the probability of significant changes of wind and solar resources due to global warming in South Africa to be small. 
Jerez et al. \cite{jerez2015impact} find, studying EURO-Cordex data, changes of solar potentials due to climate change in the range between -14\% and 2\% with largest decreases in Northern Europe.
Schaeffli et al. \cite{schaefli2007climate} incorporate a large number of climate change scenarios and find that potential climate change has a statistically significant negative effect on a hydropower plant in the Swiss Alps.
Bergstrom et al. \cite{bergstrom2001climate} study the impact of climate change on hydro resources in Sweden using different global circulation models and find that the yearly runoff circles are likely to change significantly. The global water model WaterGAP is used by Lehner et al. \cite{lehner2005impact} to study the impact of climate change on hydro potentials finding unstable regional trends in hydropower potentials with reductions of 25\% and more.
Francois et al. \cite{franccois2017effects} conclude that climate change causes temperatures and precipitation to rise in Norway, which in turn has a positive effect on the availability of surface runoff and therefore hydro power.\\ 
Other works have partially studied effects of climate change on a European power system:
Wohland et al. study the effect of climate change on a European power system \cite{wohlandmore}.
However, their study focuses on wind alone, uses a simplified version of a European power system without cost optimisation and different metrics to study effects of climate changes on renewable resources. They find more homogeneous wind conditions over Europe resulting in intensified simultaneous generation shortfalls.
A similar study is conducted by Weber et al. \cite{weber2017impact} finding that changes in temporal variability of renewables resources such as an increase of seasonal variability due to climate change likely increase the future need for backup energy and/or storage.
Kozarcanin et al. \cite{kozarcanin2018climate} use key metrics such as short-term variability or dispatchable energy to quantify the impacts of climate change on a renewable power system. They conclude that different climate change scenarios have minor effect on these key metrics.

In this work, we study the effects of climate change on renewable resources and how changes in their availability effect the cost-optimal design of a highly renewable European power system and present different novel results:
Based on a detailed model of a pan-European power system, we
quantify the effect of strong climate change for different global circulation models on the cost-optimal mix and distribution of technologies. This allows modellers and policy makers to draw direct conclusions.
By comparison with modelling based on historical weather data, we are able to quantify additional costs to the power system based on misallocation arising from using this historical weather data.

This paper is organised as follows: in Sec. \ref{sec:data} we describe the data and in Sec. \ref{sec:met} the methodology.
Sec. \ref{sec:res} discusses climate change effects on renewable resources and in
Sec. \ref{sec:costoptimal} the renewable European power system optimisation with respect to costs is analysed. Sec \ref{sec:discussion} compares the findings with other studies and discusses limitations.
Sec. \ref{sec:conclusion} concludes the study.

\begin{multicols}{2}
\nomenclature[C]{$z_0$}{surface roughness length}
\nomenclature[C]{$z$}{height above ground level}
\nomenclature[C]{$u(z)$}{wind speed at height $z$ above ground level}
\nomenclature[C]{$\text{TTI}$}{total tilted irradiation}
\nomenclature[C]{$\text{TDI}$}{tilted direct irradiation}
\nomenclature[C]{$\text{GTI}$}{ground tilted irradiation}
\nomenclature[C]{$\text{DTI}$}{diffuse tilted irradiation}
\nomenclature[C]{$\eta$, $\eta_{\text{inv}}$, $\eta_{\text{ref}}$}{solar panel related efficiencies}
\nomenclature[C]{$A$, $B$, $C$, $D$, $E$, $F$}{solar panel parameters}
\nomenclature[C]{$T_{\text{amb}}$, $T_{\text{std}}$}{ambient air temperature and standard reference temperature}
\nomenclature[C]{$\bar{g}^{\text{PV}}$}{solar panel PV generation}
\nomenclature[C]{$m$}{mass of surface runoff}
\nomenclature[C]{$g$}{gravitational constant on Earth}
\nomenclature[C]{$h$}{height above sea level}
\nomenclature[C]{$U$}{gravitational potential}
\nomenclature[C]{$\text{d}A$}{differential area element}
\nomenclature[C]{$f$}{hydro generation normalisation constant}
\nomenclature[C]{$t$}{time index}
\nomenclature[C]{$n$}{node index}
\nomenclature[C]{$I^{\text{H}}_n(t)$, $\left<I^{\text{H}}_n(t)\right>$}{inflow into hydro reservoirs and run-off-river plants at a time index $t$ and the corresponding average hourly inflow}
\nomenclature[C]{$\left<G^{\text{H}}_n(t)\right>$}{today's average hourly generation from hydro plants}

\nomenclature[R]{$\bar{g}$}{generation from an arbitrary energy carrier}
\nomenclature[R]{$g_{\text{nom}}$}{nominal generation from an arbitrary energy carrier}
\nomenclature[R]{$\text{cap}$}{capacity factor of an arbitrary energy carrier}
\nomenclature[R]{$x_{ij}$}{coordinates of a grid cell within the EURO-CORDEX domain}
\nomenclature[R]{$\tau$}{kind of renewable resource}
\nomenclature[R]{$\text{cov}(\tau(x_{ij}),\tau(x_{mn}))$}{covariance of the resource $\tau$ for the grid cell pair $x_{ij}$ and $x_{mn}$}
\nomenclature[R]{$\sigma_{\tau(x_{ij})}$}{standard deviation of the resource $\tau$ at the grid cell $x_{ij}$}
\nomenclature[R]{$\rho^{\tau}(x_{ij},x_{mn})$}{Pearsson correlation coefficient among the grid cells $x_{ij}$ and $x_{mn}$}
\nomenclature[R]{$a^{\tau}_{ij}$, $\left(a^{\tau}_{ij}\right)^{-1}$}{inverse correlation length and correlation length at the grid cell indicated by the indices $ij$}
\nomenclature[R]{$d(x_{ij},x_{mn})$}{distance between the pair of grid cells $x_{ij}$ and $x_{mn}$ on an oblate ellipsoid}
\nomenclature[R]{$\epsilon$}{ordinate interception of the correlation fit function}

\nomenclature[O]{$c_{n,s}$}{investment costs of generators of technology $s$ at node $n$}
\nomenclature[O]{$G_{n,s}$}{capacities of generators of technology $s$ at node $n$}
\nomenclature[O]{$c_l$}{investment costs of transmission capacities at link $l$}
\nomenclature[O]{$F_l$}{transmission capacities of link $l$}
\nomenclature[O]{$o_{n,s}$}{marginal costs of generation of technology $s$ at node $n$}
\nomenclature[O]{$g_{n,s,t}$}{dispatch of generators of technology $s$ at node $n$ and time $t$}
\nomenclature[O]{$d_{n,t}$}{demand at node $n$ and time $t$}
\nomenclature[O]{$K_{n,l}$}{incidence matrix of the network}
\nomenclature[O]{$f_{l,t}$}{flows over link $l$ at time $t$}
\nomenclature[O]{$g^-_{n,s,t}$}{minimal dispatch of generators or storage units of technology $s$ at node $n$ and time $t$}
\nomenclature[O]{$\bar{g}_{n,s,t}$}{maximal dispatch of generators or storage units of technology $s$ at node $n$ and time $t$}
\nomenclature[O]{$\eta_{0,s}$, $\eta_{1,s}$, $\eta_{2,s}^{-1}$}{standing losses, charging and discharging efficiencies of storages of technology $s$}
\nomenclature[O]{$soc_{n,s,t}$}{state of charge of storage of technology $s$ at node $n$ and time $t$}
\nomenclature[O]{$\text{inflow}_{n,s,t}$}{storage inflow of technology $s$ at node $n$ and time $t$}
\nomenclature[O]{$\text{spillage}_{n,s,t}$}{storage spillage of technology $s$ at node $n$ and time $t$}
\nomenclature[O]{$L_l$}{length of link $l$}  
\nomenclature[O]{$\text{CAP}_F$}{global limit of the sum of all single transmission line capacities}
\nomenclature[O]{$\text{CAP}_{\text{CO}_2}$}{global limit on $\text{CO}_2$ emissions}  
\nomenclature[O]{$\eta_{n,s}$}{efficiencies of generators of technology $s$ at node $n$}  
\nomenclature[O]{$e_{n,s}$}{$\text{CO}_2$ emissions of generators of technology $s$ at node $n$}
\nomenclature[O]{$s$}{technology}
\nomenclature[O]{$n$}{node index}
\nomenclature[O]{$l$}{link index}
\nomenclature[O]{$t$}{time index}
\printnomenclature
\end{multicols}

\section{Data}
\label{sec:data}
Renewable energy production crucially depends on the availability of renewable resources.
Wind turbines convert the kinetic energy of wind to electric power, whereas solar photovoltaic (PV) panels convert irradiation into electricity using semiconductor materials that exhibit the photoelectric effect.
Hydro power, which is the third major renewable energy source in terms of globally installed generation capacity \cite{irena2018}, uses the potential energy of dammed or running water to drive turbines and generators.
A common method to calculate the energy inflow into hydro dams and run-off-river plants is to make use of surface water runoff from meteorological reanalysis models \cite{kies2016effect}. We adopt this method and look at the availability of potential energy from total water runoff to characterise the potentials of hydro power.

\subsection{Climate Projection Data}
We use three ensemble members of the EURO-CORDEX project \cite{jacob2014euro} as climate change affected weather datasets. 
Other available models providing climate change affected weather data are for instance the PRUDENCE \cite{christensen2007summary} and ENSEMBLES \cite{hewitt2005ensembles, hewitt2009towards} for Europe and NARCAP \cite{mearns2005narccap} for the USA.
Each ensemble member is based upon a unique general circulation model (GCM), which is a global numerical climate model on a coarse spatial grid and well suited to replicate large-scale circulation features of the climate \cite{wilby2004guidelines}. However, they are not accurate on the regional scale. To remedy this problem, GCM data are used to drive higher spatially resolved regional climate models (RCM).
The downscaling to a finer spatial resolution within the EURO-CORDEX project was performed with a single regional climate model, the RCA-4 model from the SMHI \cite{strandberg2015cordex}.
The different ensemble members as well as the acronym that is assigned to each member throughout this study are shown in Table \ref{table:ensemble_members}. We focus on these three members, because they rely on the same RCM as well as calendaric system (Gregorian Calendar). For some of the following investigations, we split the weather data time series into four distinct periods as given in Table \ref{table:periods}.

\begin{table}[!t]
\centering
\begin{tabular}{||c|c||}
\hline
\textbf{Acronym} & \textbf{EURO-CORDEX model}\\
\hline

CNRM & CNRM-CERFACS-CNRM-CM5\\
ICHEC & ICHEC-EC-EARTH\\
MPI & MPI-M-MPI-ESM-LR\\
\hline
\end{tabular}
\caption{EURO-CORDEX ensemble members considered.}
\label{table:ensemble_members}
\end{table}

Each member provides a dataset containing the weather data variables near-surface wind speed (sfcWind), surface downwelling and surface upwelling shortwave radiation (rsds, rsus), and near-surface ambient air temperature (tas), as well as total runoff (mrro). The parenthesised terms describe the CORDEX variable standard names. Most variables are provided with a spatial resolution of 0.11\degree  and a temporal resolution of 3 hours.

We focus on the representative concentration pathway (RCP) 8.5 scenario \cite{stocker2013climate}, which is characterised by an increase in radiative forcing of 8.5 W/m$^2$ around the year 2100 relative to pre-industrial values \cite{weyant2009report}. It is the most extreme scenario. The likely range of global average temperature increase in the period 2081-2100 associated with this scenario amounts to 2.6 to 4.8\degree C in comparison to pre-industrial times. RCP scenarios were used in the last three assessment reports of the IPCC \cite{stocker2013climate,mccarthy2001climate,change2007fourth} as well.

\subsection{Electricity Consumption Data}
Electricity consumption data used in this work originate from the European Network of Transmission System Operators for Electricity (ENTSO-E) (\cite{ENTSOE}) as the historical time series of the year 2012, which is, for the considered European countries, applied to each year until 2100. The overall electricity demand of all countries sums up to around 3000 TWh/a and consists solely of electric energy demand. The transport sector and thermal heating are not included. Heating and transport play an important role in Europe's energy consumption: however, including them would not have a significant impact in results, since increasing the demand would mostly increase overall numbers of installed capacities.
It furthermore is expected that climate change in Europe  does not only affect the overall demand for electricity, but also its temporal patterns in an inhomogeneous way (\cite{wenz2017north}, \cite{bossmann2015shape}); for the same reasons, this aspect has been neglected, too.
However, it would be a straightforward extension to include climate-affected forecast load data.

\begin{table}[!t]
\centering
\begin{tabular}{||c|c|c||}
\hline
\textbf{Acronym} & \textbf{Period} &\textbf{Years covered}\\
\hline
HIS & Historical Epoch & 1970 - 2005\\
BOC & Begin of Century & 2006 - 2037\\
MOC & Mid of Century & 2038 - 2069\\
EOC & End of Century & 2070 - 2100\\
\hline
\end{tabular}
\caption{Time periods considered.}
\label{table:periods}
\end{table}

\section{Methodology}
\label{sec:met}

The sole fossil generation source incorporated in the model is gas, represented by open cycle gas turbines (OCGT). 
The major reason for the choice to only allow gas turbines as fossil power source are the strict long term decarbonisation goals, which contradict high emission technologies. Gas-fired turbines are often considered a low-cost option technology to integrate renewables in highly decarbonised power systems \cite{brouwer2016least}.
Besides OCGT and the different types of renewable power sources, i.e. onshore wind, offshore wind, PV and hydro, the model incorporates transmission lines and different types of energy storage units, i.e., hydrogen storage, batteries, and pumped hydroelectric storage (PHS).

\subsection{Conversion of Renewable Resources to Power}
Weather data is converted to potential power output from renewable power plants for each grid cell using the Aarhus Renewable Atlas \cite{andresen2015validation}.\\

\begin{figure}
\centering
\includegraphics[width=.4\textwidth]{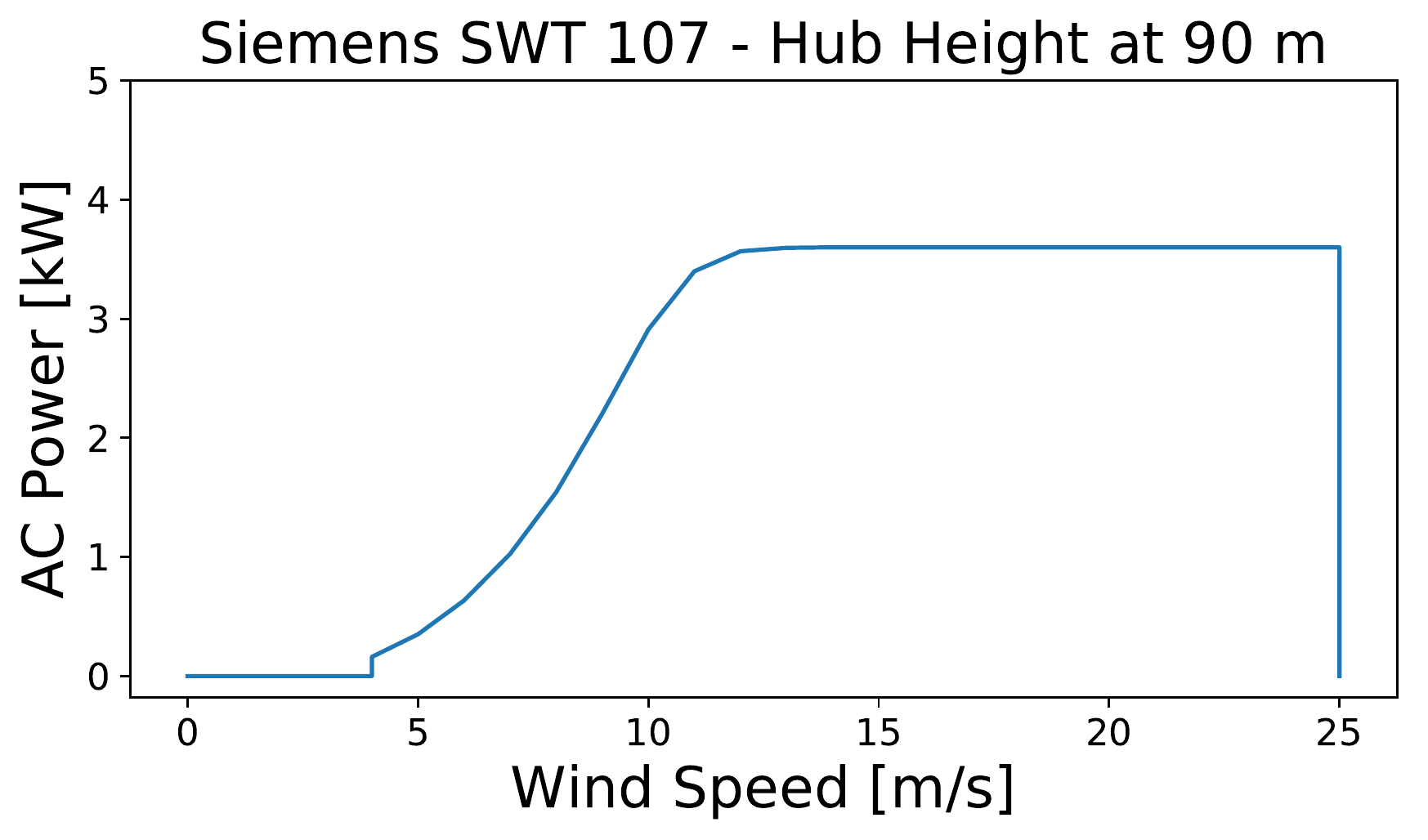}     \includegraphics[width=.4\textwidth]{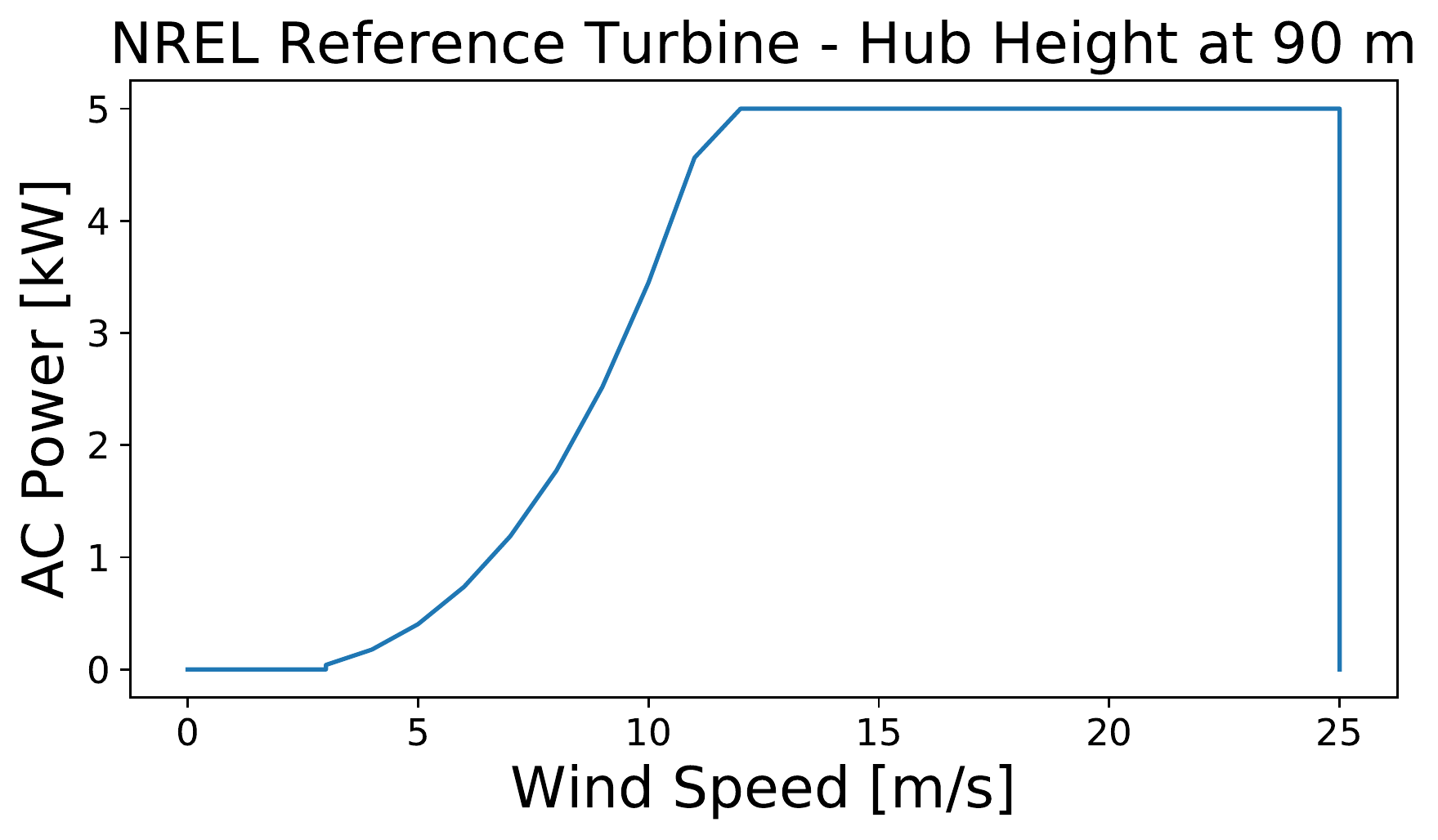}
\caption{Wind turbine power curves that were used to convert wind speed at hub height to power. Left: Siemens SWT 107 for onshore locations, right: NREL Reference Turbine for offshore locations.}
\label{fig:power_curves_wind}
\end{figure}

\underline{Wind Turbines}\\
Near-surface wind speeds are upscaled to the wind turbine's hub height under
the assumption of  logarithmic wind profiles. The logarithmic wind profile is a semi-empirical relationship between height and wind speed in the lowest 100 m of the troposphere and valid under neutral stability conditions. It is given by
\begin{align*}
    \frac{u(z_2)}{u(z_1)} &= \frac{\log(\frac{z_2}{z_0})}{\log(\frac{z_1}{z_0})},
\end{align*}
where $u$ is the wind speed, $z_2$ is the chosen hub height (90 m) and $z_1$ is the height at which wind speeds are given (10m). $z_0$ is the surface roughness length, which is provided by the datasets as a static quantity.
Wind speeds at hub height are then converted to power using the power curve of a Siemens SWT 107 turbine for onshore locations and the power curve of a NREL Reference Turbine for offshore locations. Both power curves are depicted in Fig. \ref{fig:power_curves_wind}.
The power curves are additionally smoothed with a Gaussian kernel to better match actual wind feed-in data  \cite{andresen2015validation}.
Although wind power turbines of the future might change considerably (e.g., different hub heights or multi-rotor instead of single rotor), properties of conversion from wind to power are strongly determined by laws of physics and the results are most affected by the cost of a certain electricity type. However, our study focuses on the influence of climate change alone and does not investigate this aspect further. \\

\underline{Solar Panels}\\
For PV power, several steps are performed: first, rsds fields are split retrospectively into direct and diffuse horizontal irradiation using the Reindl Clearsky model \cite{reindl1990evaluation} and thereupon rotated on a tilted surface. The tilted rotated irradiation consists of three parts:

First, the tilted direct irradiation (TDI) is simply the geometrically rotated direct horizontal irradiation.
The diffuse tilted irradiation (DTI) is the rotated diffuse irradiation, where the rotation is executed subsequent to the Hay-Davies model \cite{daviescalculation}.
The ground tilted irradiation (GTI) is determined from the ground albedo invoking the rsus fields.
The total irradiation on the tilted plane (TTI) is finally given as the sum of its components:
\begin{align}
    \text{TTI} &= \text{TDI} + \text{GTI} + \text{DTI}.
\end{align}

Potential power generation from PV is then calculated using an effective solar panel model from Beyer et al. \cite{beyer2004robust} as a function of total tilted irradiation:
\begin{align}
    \bar{g}^{\text{PV}} &= \text{TTI} \cdot \eta \cdot \eta_{\text{inv}}
\end{align}
The efficiency $\eta$ is given by
\begin{align}
    \eta &= \eta_\text{ref} \cdot \frac{1+D\left(E \cdot \text{TTI} + \Delta T\right)}{1 + D\cdot F \cdot \text{TTI} \cdot \eta_\text{ref}}
\end{align}
with
\begin{align}
    \eta_{\text{ref}} &= A + B \cdot \text{TTI} + C \cdot \log(\text{TTI}),
\end{align}
where $\Delta T = T_\text{amb} - T_\text{std}$ describes the difference between ambient air and standard test temperature (298 K). The quantities $A$ to $F$ are device-specific parameters and for the chosen Kanena hybrid panel given by $A = 0.066$, $B = -4.443 \cdot 10^{-6}$, $C = 0.012$, $D = -0.0041 /$K, $E = 0.031$ Km$^2$/W and $F = 0.034$ Km$^2$/W.
The inverter efficiency $\eta_\text{inv}$ is chosen to be 0.9.

From detailed studies on the estimations concerning average capacity factors of solar PV made by Pietzcker et al. \cite{pietzcker2014} for the European countries, we find that the power output from solar panels described by the given methodology is, averaged over these European countries, overestimated by approximately 21\%. 
For this reason, all hourly PV capacity factors were reduced by this factor. 
It should be noted that overestimation of radiation is a well-known issue of climate models \cite{wild2008short}.\\

\underline{Hydro Plants}\\
To model hydro power, we use a potential energy approach using water runoff data.
The potential gravitational energy of a mass $m$ relative to the sea level is given by
\begin{align}
U &= m \cdot g \cdot h,
\end{align}
where $g = 9.81 \ \text{m}/\text{s}^2$ is the gravitational acceleration on Earth and $h$ the height above sea level.
For each grid cell spanning an area $\text{d}A$ in a given country, the inflow into hydro reservoirs and run-off-river plants is calculated as a linear function of the potential energy of the water runoff variable:
\begin{align}
 I^{\text{H}}_n(t) &=  f \cdot g \int_A m(x,y,t) \ h(x,y) \ \text{d}A
\end{align}
$f$ is a normalization constant that ensures $\left<I^{\text{H}}_n(t)\right>$ = $\left<G^{\text{H}}_n(t)\right>$, where 
the latter is today's average hourly generation from hydro plants in the corresponding country. 
Normalisation data for each country is taken from the RESTORE 2050 project \cite{kies2016restore} and the EIA \cite{eia2011international}, covering the years 2000 until 2014.

\subsection{Renewable Energy Resource Analysis}
To study the effect of global warming on renewable resources we investigate regionally resolved changes in capacity factors of several renewable energy carrier types and also correlation lengths of some of the underlying weather data variables.\\

\underline{Capacity Factors}\\
The capacity factor of a power plant is given by the mean power generation in a considered temporal period over its nominal nameplate capacity:
\begin{equation}
\mathrm{cap} = \frac{\bar{g}}{g_{\text{nom}}}
\end{equation}
The capacity factors were computed for the periods from Table \ref{table:periods} for onshore wind,  offshore wind as well as PV.\\

\underline{Correlation Lengths}\\
The correlation matrix is given by the Pearsson correlation coefficient for pairs of grid cells ($x_{ij},x_{mn}$):
\begin{equation}
\rho^\tau\left(x_{ij},x_{mn}\right) = \frac{\mathrm{cov}(\tau(x_{ij}),\tau(x_{mn}))}{\sigma_{\tau(x_{ij})} \cdot \sigma_{\tau(x_{mn})}},
\end{equation}
where $\mathrm{cov}$ denotes the covariance, $\sigma$ the standard deviation and $\tau$ the renewable resource time series from the climate projection's weather data.

\begin{figure}
\begin{center}
\includegraphics[width=.3\textwidth]{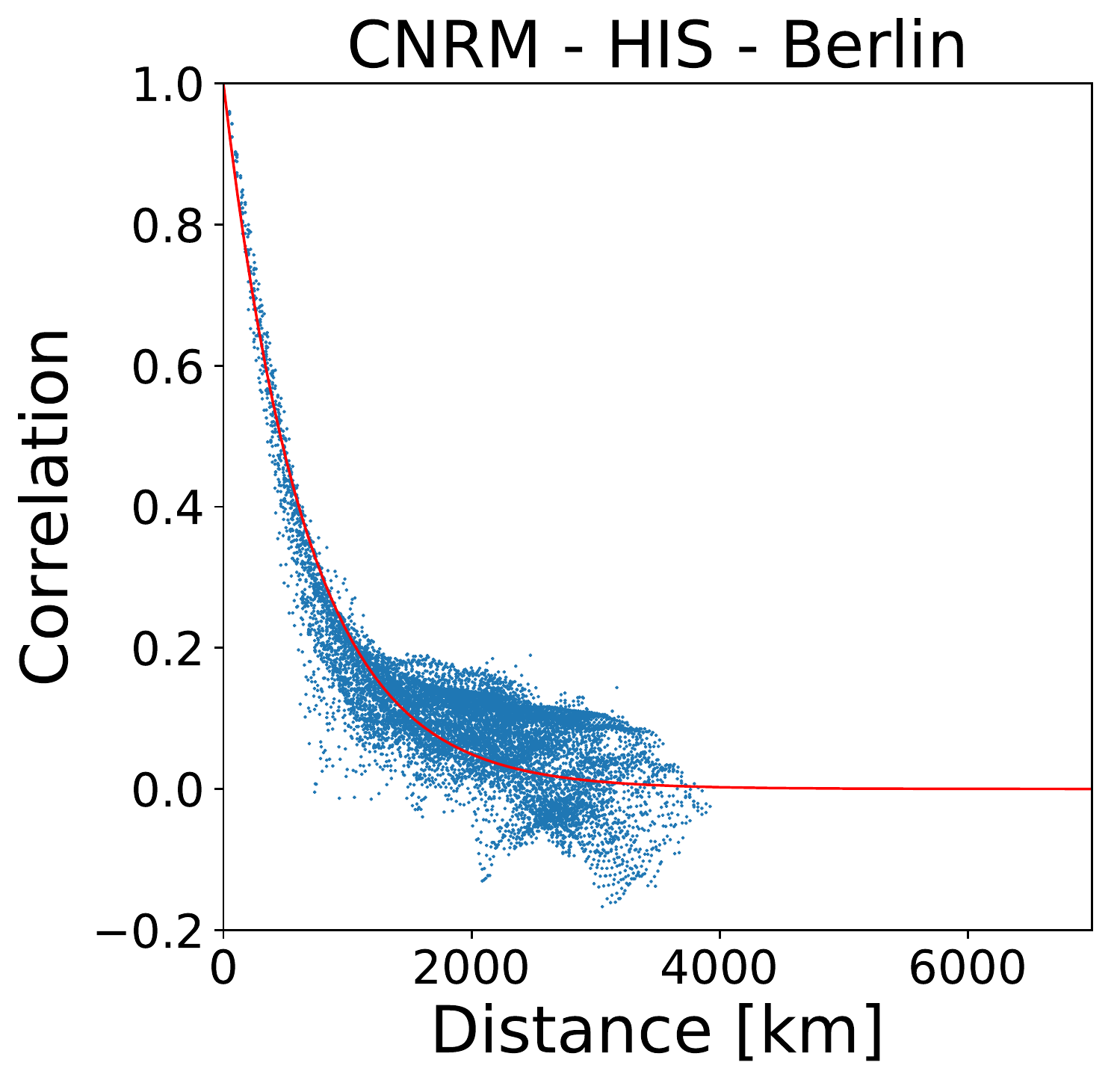}
\includegraphics[width=.3\textwidth]{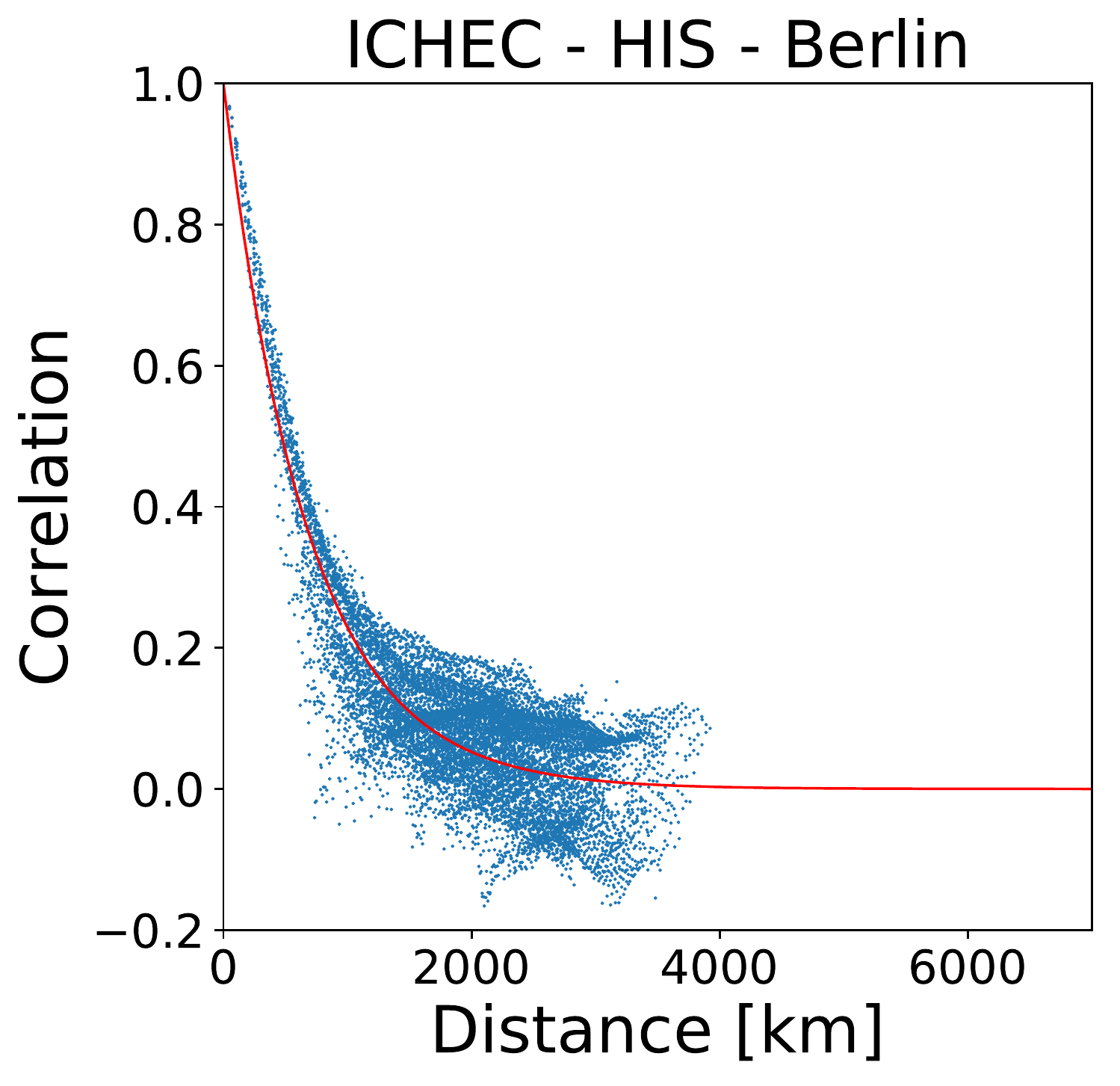}
\includegraphics[width=.3\textwidth]{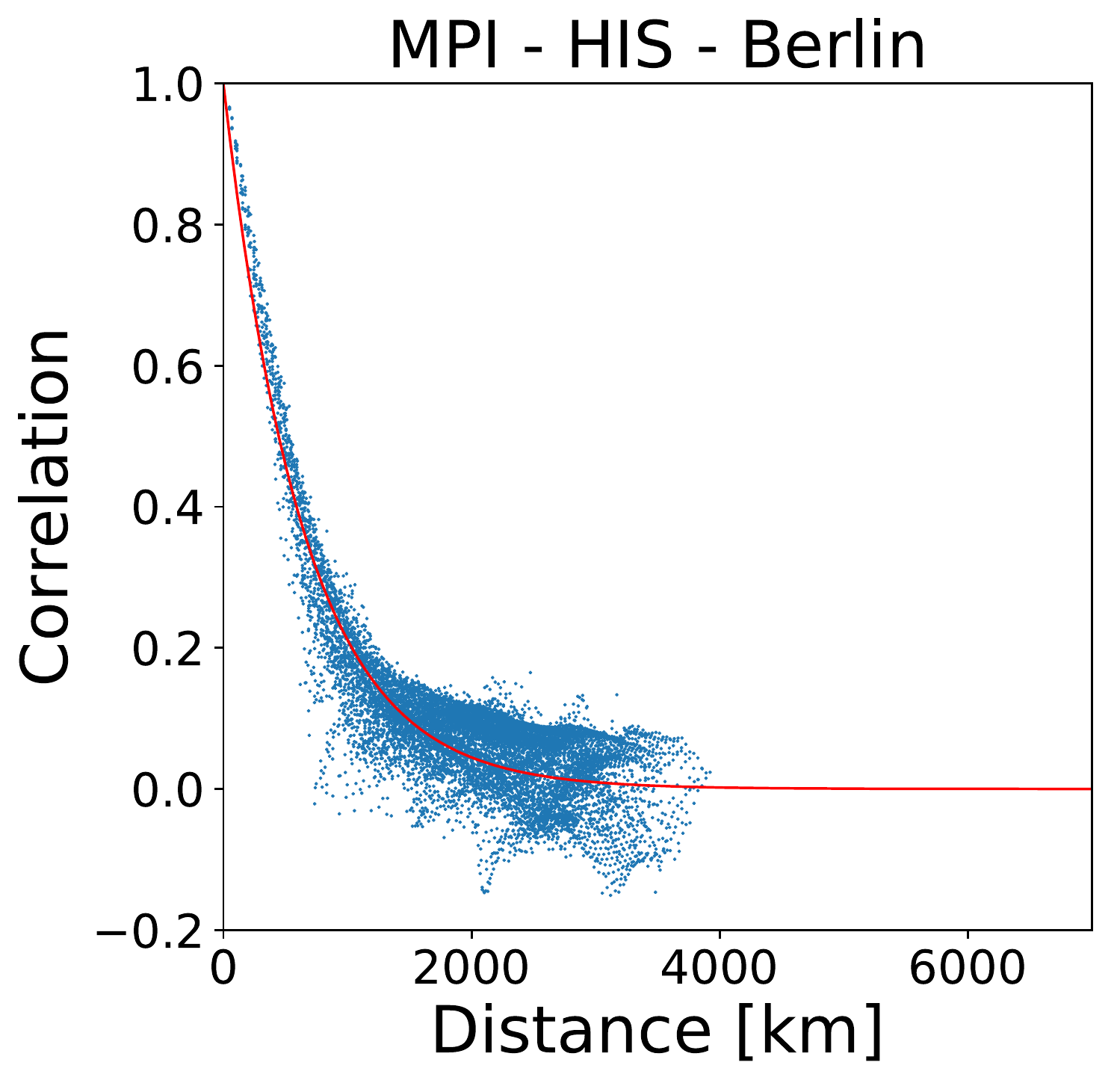}
\caption{\label{fig:fits}  Correlation coefficients over distance between the grid cell containing Berlin (lat=52.54\degree, lon=13.58\degree) and all other cells on the grid for the renewable resource wind.}
\end{center}
\end{figure}

The relationship between the correlation coefficient for wind speeds and distance then was assumed to be of an exponential form following the conclusion of Hasche \cite{hasche2010general}.
The expression
\begin{equation}
 \rho^\tau\left(x_{ij},x_{mn}\right) = \exp\left(- a^\tau_{ij} \cdot d\left(x_{ij},x_{mn}\right)\right) + \epsilon,
\end{equation}
where $d$ denotes the distance between a pair of grid cells, was fitted for each single grid cell using least squares. The resulting inverse fit parameter $\left(a_{ij}^{\tau}\right)^{-1}$ is interpreted as the local correlation length.
Correlation lengths were computed for all four periods from Table \ref{table:periods} for wind speeds. Exemplary correlation length fits for the grid cell that includes Berlin are shown in Fig. \ref{fig:fits}. A good agreement with the assumption of exponentially decreasing correlation coefficients is observed.


\subsection{Power System Optimisation}
\label{ss:psopt}

Changes in the availability of renewable resources have a direct influence on renewable power systems; therefore, we choose to examine the optimisation of a highly renewable European power system with respect to economic costs.
The European power system studied in this work is simplified in such a way that each country is represented by a single node and countries are interconnected by links via transmission lines representing today's topology. Within single countries, renewable and conventional generation facilities as well as storage units are allowed to be built.
This methodology has been used in a variety of papers, e.g., by Schlachtberger et al. to study the benefits of cross-country cooperation \cite{schlachtberger2017benefits}, by Rodriguez et al \cite{rodriguez2014transmission} and Becker et al. \cite{becker2014transmission} to study the need for inter-connecting transmission capacities, by Kies et al. \cite{kies2016demand} to study the European demand side potential, by Heide et al. \cite{heide2010seasonal} to study seasonal balancing effects of renewables and by Brown et al. \cite{brown2018synergies} to study synergies of transmission grid extensions and sector coupling.
Setup, methodology and cost assumptions are mainly adopted from Schlachtberger et al. \cite{schlachtberger2017benefits}. 
We invoke the same model, but use different weather data, i.e. the EURO-CORDEX climate projection data from the models described in Table \ref{table:ensemble_members}. 

The power system is formulated as a linear optimisation model that minimises total system costs. 
In the following, a short general overview on the methodology is given. To solve the system of linear equations occurring under the optimisation, we use the software toolbox Python for Power System Analysis \cite{brown2017pypsa}.\\

\begin{table*}[!t]
\begin{center}
\resizebox{.8\textwidth}{!}{\begin{tabular}{ lrrrrrr }

\hline
Technology & Capital Cost & Marginal cost & Efficiency \\ 
           &  [Euro/GW/a] & [Euro/MWh] & dispatch/uptake \\ 
\hline
OCGT & 47,235 & 58.385 & 0.390 \\
Onshore Wind & 136,428 & 0.015 & 1\\
Offshore Wind & 295,041 & 0.020 & 1 \\
PV & 76,486 & 0.010 & 1 \\
Run-Off-River & 0 & 0 & 1 \\
Hydro Reservoir & 0 & 0 & 1 / 1\\
PHS & 0 & 0 & 0,866 / 0,866\\
Hydrogen Storage & 195,363 & 0 & 0.580 / 0.750\\
Battery & 120,389  & 0 & 0.900 / 0.900\\
\hline
Transmission Lines & 16,450 & mEuro/TWkm/a \\
\hline


\end{tabular}}
\caption{Annualised cost assumptions for generation and storage technologies as well as transmission lines, originally based on 2030 value estimates from Schroeder et al. \cite{schroeder2013current}.}
\label{tab:costsassumptions}
\end{center}
\end{table*} 

\underline{Objective Function}\\
The optimisation objective is the minimisation of the total system costs, where the objective function reads
\begin{equation} \label{eq:main}
\min_{g,G,f,F} \left(\sum_{n,s} c_{n,s} \cdot G_{n,s} + \sum_l c_l \cdot F_l + \sum_{n,s,t} o_{n,s} \cdot g_{n,s,t}\right)
\end{equation}
Here, $c_{n,s}$ are the investment costs for generation capacity, $c_l$ are the investment costs for transmission line capacity, $o_{n,s}$ are the marginal costs of energy generation, $G_{n,s}$ and $F_l$ are the capacities of generators and transmission lines and $g_{n,s,t}$ are the time series of generation.
The index $n$ runs over all discrete nodes of the network, $l$ over all inter-connecting links and $s$ over all considered energy technologies, e.g. onshore wind, but also storage units like batteries. A list of all energy carrier types used, the corresponding modelling details as well as their cost assumptions are given in Table \ref{tab:costsassumptions}.\\

\underline{Constraints}\\
In addition to the objective, multiple constraints have to be satisfied.
To ensure a stable power system operation, the demand must be matched by the actual power generation, transmission and/or storage flows in space and time:
\begin{equation}
\sum_s g_{n,s,t} - d_{n,t} = \sum_l K_{n,l} \cdot f_{l,t}\quad \forall \quad n,t
\end{equation}
Here $d_{n}$ is the demand, $K$ is the incidence matrix of the network that describes its topology, and $f_l$ is the flow through a link $l$. The distribution of renewable generator types within the nodes was chosen proportional to the site quality of each carrier type, i.e. proportional to its corresponding capacity factors.

The dispatch $g_{n,s,t}$ of a generator or storage unit within a node is constrained by its maximal capacity $G_{n,s}$ multiplied by the corresponding hourly capacity factor $\bar{g}_{n,s,t}$:
\begin{equation}
{g}^-_{n,s,t} \cdot G_{n,s} \leq g_{n,s,t} \leq \bar{g}_{n,s,t} \cdot G_{n,s} \quad \forall \quad n,t
\end{equation}
The maximal capacity of the renewable generator types was determined in accordance with Hoersch/Brown et al. \cite{horschpypsa} invoking the Corine and Natura2000 land use data \cite{corine2012, natura2000}. For conventional generators, dispatch is constrained by the installed capacity, so that $\bar{g}_{n,s,t} = 1.0$.

Storage consistency is ensured via the storage state of charge $\mathrm{soc_{n,t}}$ dispatch constraint:
\begin{align*}
\mathrm{soc}_{n,s,t} &= \eta_{0,s} \cdot \mathrm{soc}_{n,s,t-1} + \eta_{1,s} \cdot g_{n,s,t,\textrm{store}} -  \eta_{2,s}^{-1} \cdot g_{n,s,t,\textrm{dispatch}} \\
& \quad + \mathrm{inflow}_{n,s,t} - \mathrm{spillage}_{n,s,t} \quad \forall \quad n, s, t > 1,\\
\mathrm{soc}_{n,s,t=0} &= \mathrm{soc}_{n,s,t=t_{\text{max}}} \quad \forall \quad n, s.
\end{align*}
$\eta_0$ describes standing losses and $\eta_1$ and $\eta_2$ efficiencies of the corresponding charging and discharging processes.
The second constraint, where $t_{\text{max}}$ is the last time step, ensures a cyclic state of charge.
Line capacities are treated as free parameters and hence are unlimitedly expandable. However, if a global limit of transmission capacities is set, the total sum of line capacities can not exceed this limit:
\begin{equation}
\sum_l F_l \cdot L_l \leq \mathrm{CAP}_{F},
\end{equation}
where $L_l$ is the length of link $l$ and $\mathrm{CAP}_{F}$ the global line capacity limit. The cost-optimal power system analysed in Sec. \ref{sec:costoptimal} makes no use of these line constraints and hence transmission capacities are unlimitedly expandable.

Finally, a CO$_2$ emission constraint is enforced: 
\begin{equation}
\sum_{n,s,t} \frac{1}{\eta_{n,s}} \cdot g_{n,s,t} \cdot e_{n,s} \leq \mathrm{CAP}_{\text{CO}_2} \label{eq:co2},\\
\end{equation}
where $\eta$ denotes the technology specific combustion efficiency and $e$ the corresponding specific CO$_2$ emissions.
In all cases, CO$_2$ emissions are limited to 5\% of the electricity related emissions in the reference year 1990 in Europe ($\mathrm{CAP}_{\text{CO}_2} = 77.5 \ \text{Mt/a}$). The constraint is based on the total amount of 3.1 Gt/a CO$_2$ emissions from the sectors electricity, traffic, residentials and services. Approximately one half of this amount comes from the electricity sector and shall be reduced by 95\%.\\

\underline{Flow Equations}\\
The linearised power flows $f_l$ through each link are determined from the linearised AC power flow equations and can not exceed transmission limits:
\begin{equation}
|f_{l,t}| \leq F_l \quad \forall \quad l.
\end{equation}

\section{Renewable Energy Resources}
\label{sec:res}

\subsection{Capacity Factors}
\begin{figure}
\begin{center}
\includegraphics[width=.3\textwidth]{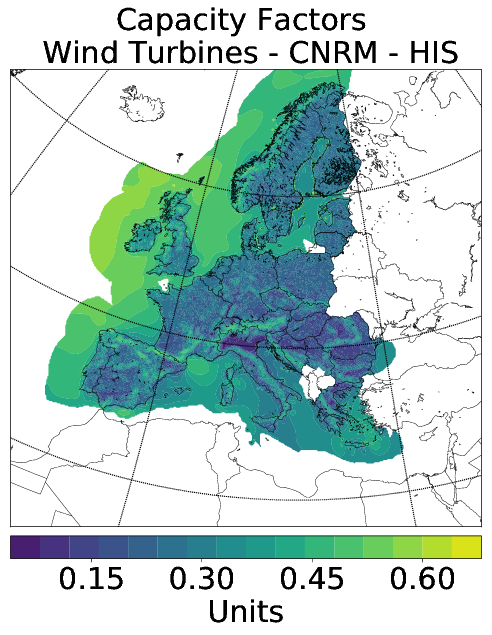}
\includegraphics[width=.3\textwidth]{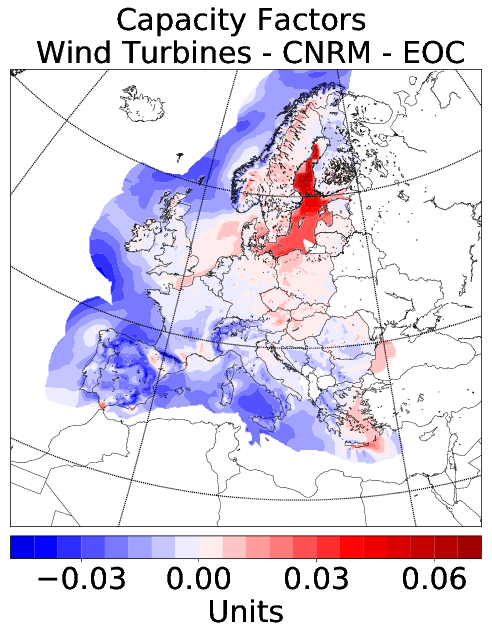}\\
\includegraphics[width=.3\textwidth]{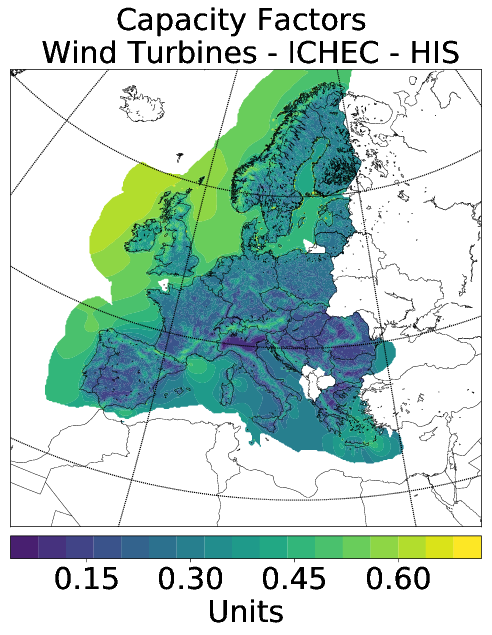}
\includegraphics[width=.3\textwidth]{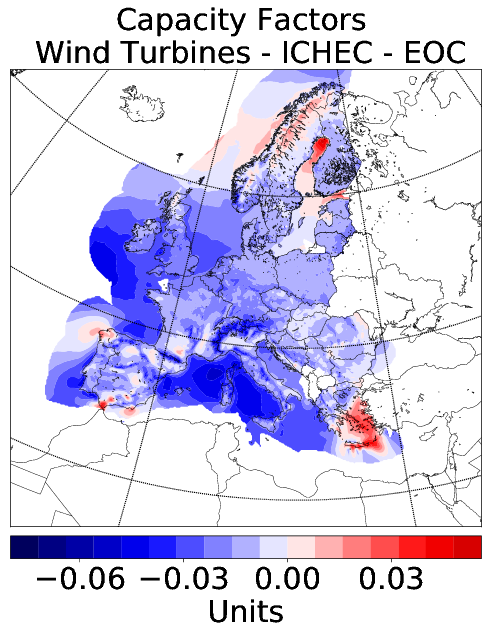}\\
\includegraphics[width=.3\textwidth]{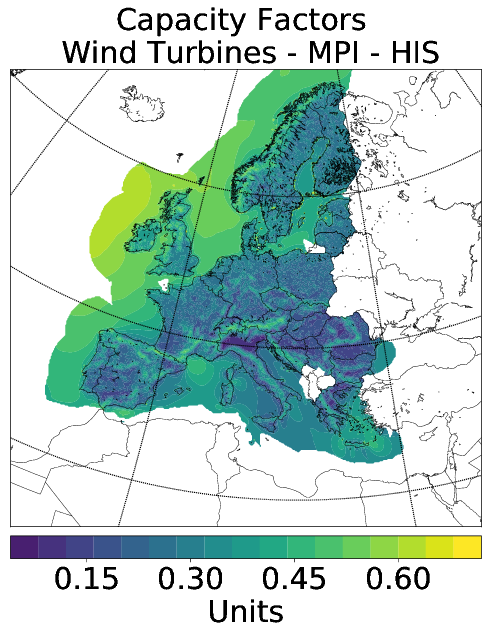}
\includegraphics[width=.3\textwidth]{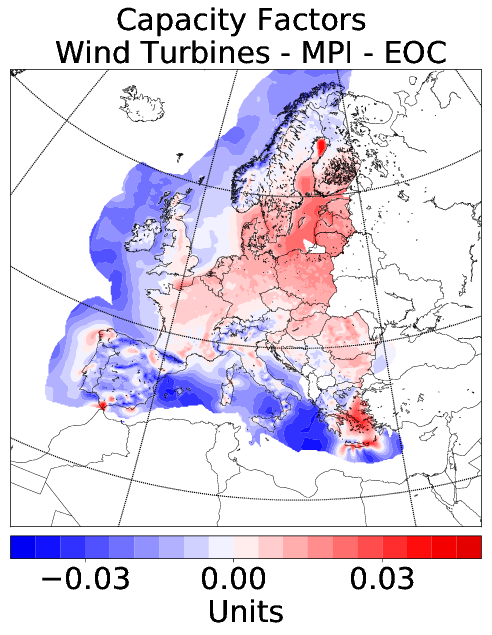}\\
\caption{\label{fig:capacity_factors_wind} Spatially resolved onshore and offshore wind generation capacity factors for the single models at the HIS period as well as the relative model changes at the EOC period compared to the corresponding HIS periods.}
\end{center}
\end{figure}

\begin{figure}
\begin{center}
\includegraphics[width=.3\textwidth]{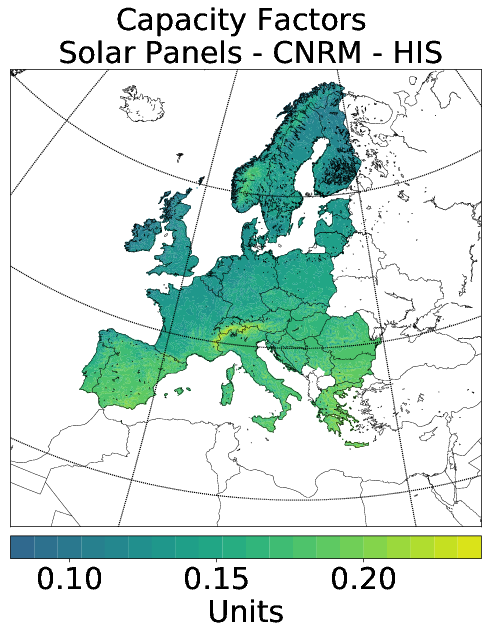}
\includegraphics[width=.3\textwidth]{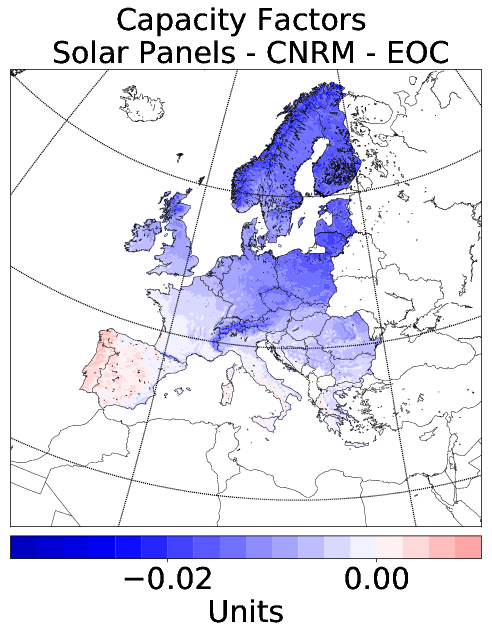}\\
\includegraphics[width=.3\textwidth]{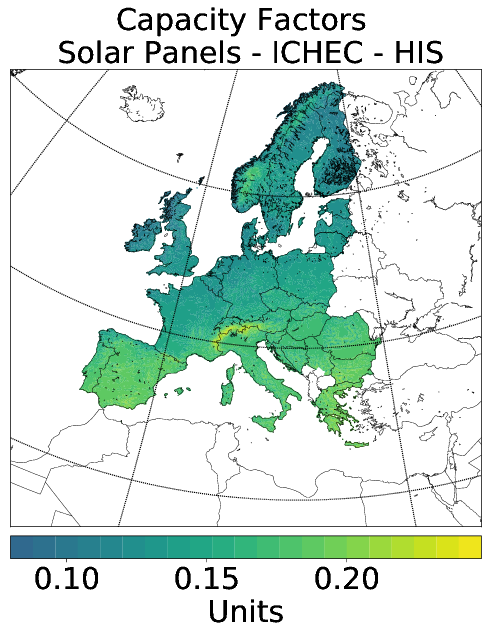}
\includegraphics[width=.3\textwidth]{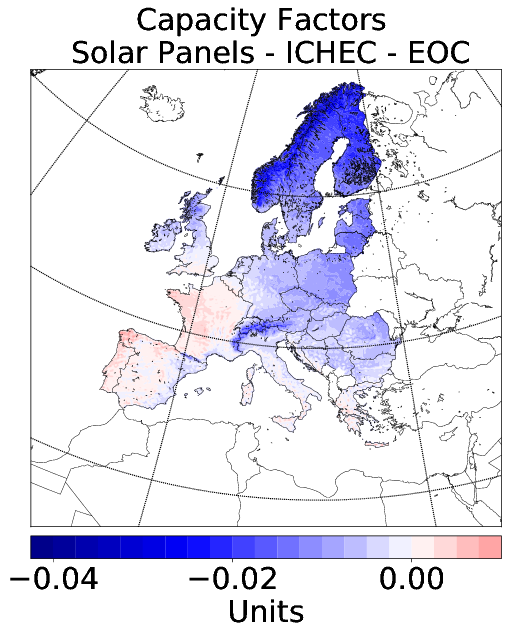}\\
\includegraphics[width=.3\textwidth]{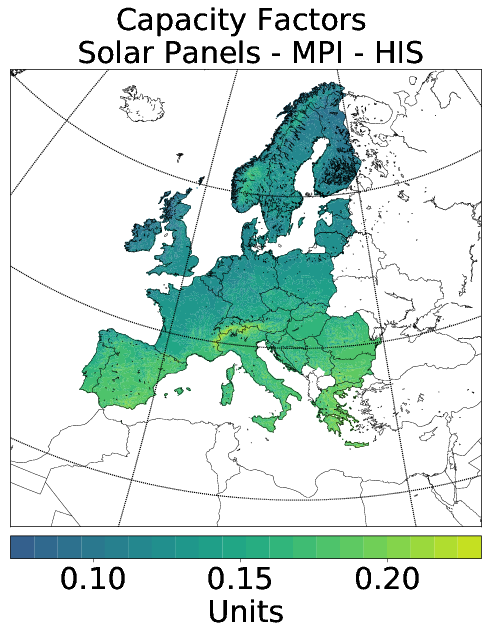}
\includegraphics[width=.3\textwidth]{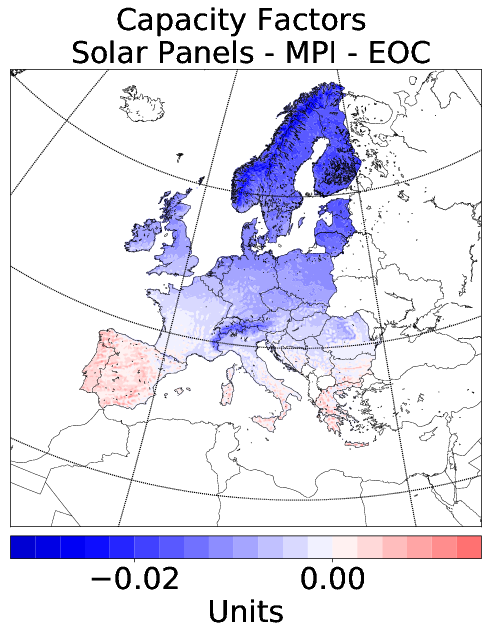}\\
\caption{\label{fig:capacity_factors_solar} Spatially resolved PV generation capacity factors for the single models at the HIS period as well as the relative model changes at the EOC period compared to the corresponding HIS periods.}
\end{center}
\end{figure}

Fig. \ref{fig:capacity_factors_wind} shows absolute values of spatially resolved capacity factors for wind power and Fig. \ref{fig:capacity_factors_solar} for solar power at the historical period and changes in the EOC period for every GCM. 
The three models show mostly similar behaviour:
Except for the Baltic sea, where results are different between models, decreasing capacity factors are observed for wind offshore.
Onshore, the MPI model predicts increasing, the CNRM model barely any changes and ICHEC model decreasing capacity factors in central Europe.
For PV, changes are small, but mostly decreasing capacity factors can be observed with the exception of Spain, parts of Italy (all models) and France (ICHEC).

\subsection{Correlation Lengths}
\begin{figure}[!htp]
\begin{center}
\includegraphics[width=.3\textwidth]{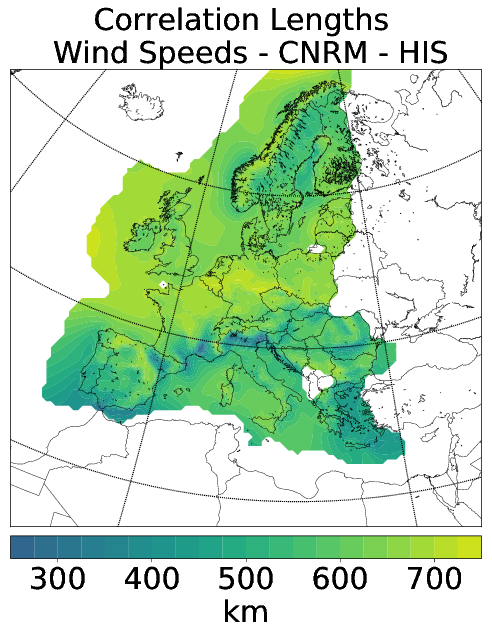}
\includegraphics[width=.3\textwidth]{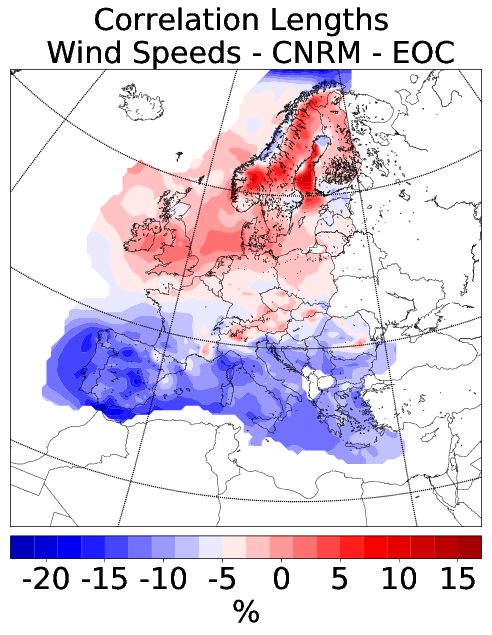}\\
\includegraphics[width=.3\textwidth]{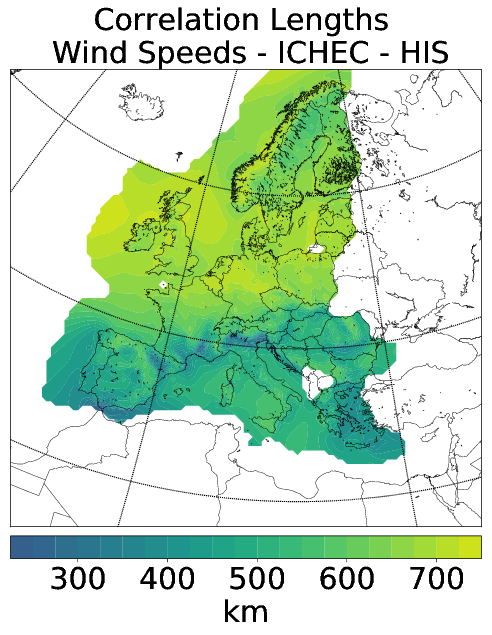}
\includegraphics[width=.3\textwidth]{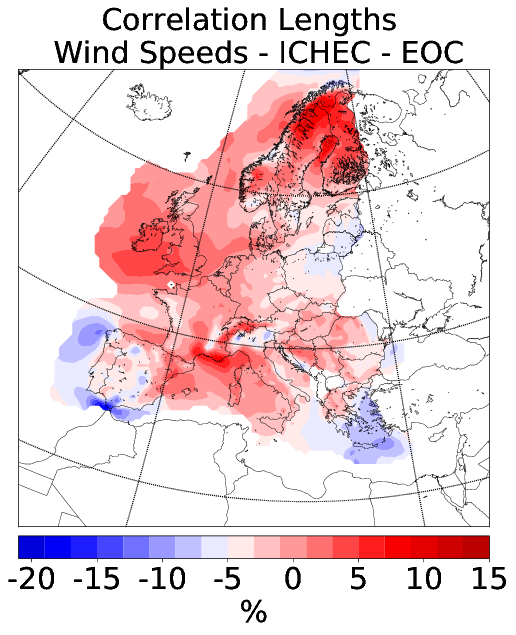}\\
\includegraphics[width=.3\textwidth]{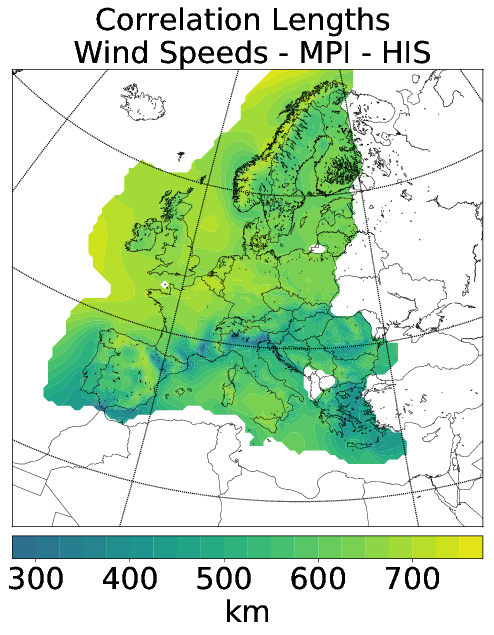}
\includegraphics[width=.3\textwidth]{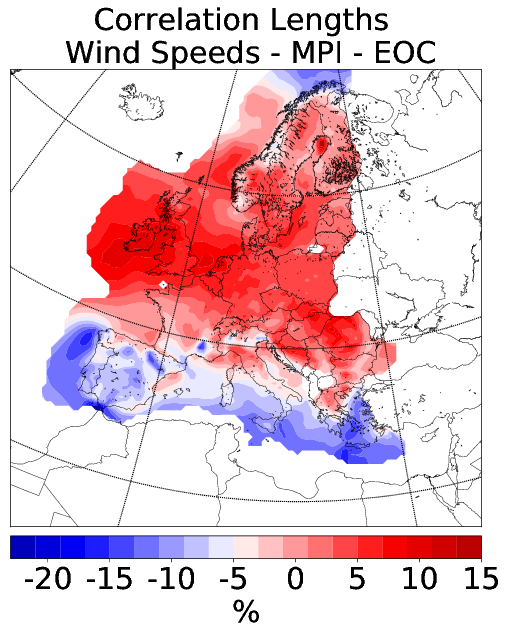}\\
\caption{\label{fig:correlation_lenghts_wind} Spatially resolved wind speed (sfcWind) correlation lengths for the single models at the HIS period as well as the relative model changes at the EOC period compared to the corresponding HIS periods.}
\end{center}
\end{figure}

Fig. \ref{fig:correlation_lenghts_wind} shows absolute values of spatially resolved correlation lengths of wind speeds at the HIS period and changes in the EOC period for the three ensemble members.
Correlation lengths are for all models in the range of 300 to 700 km for wind speeds with smaller values over irregular terrain.
Until the end of the century, the ensemble members show increasing correlation lengths by up to 15\% in most regions, with partial exceptions in the far south, i.e. in Spain, Italy and Greece.
Increasing correlation lengths for wind speeds are likely to increase the cost of wind turbine integration into a power system, as they have a negative effect on smoothing effects \cite{hasche2010general} that reduce the need for balancing measures.

\subsection{Hydro Potential}
\begin{figure}
\begin{center}

\includegraphics[width=.49\textwidth]{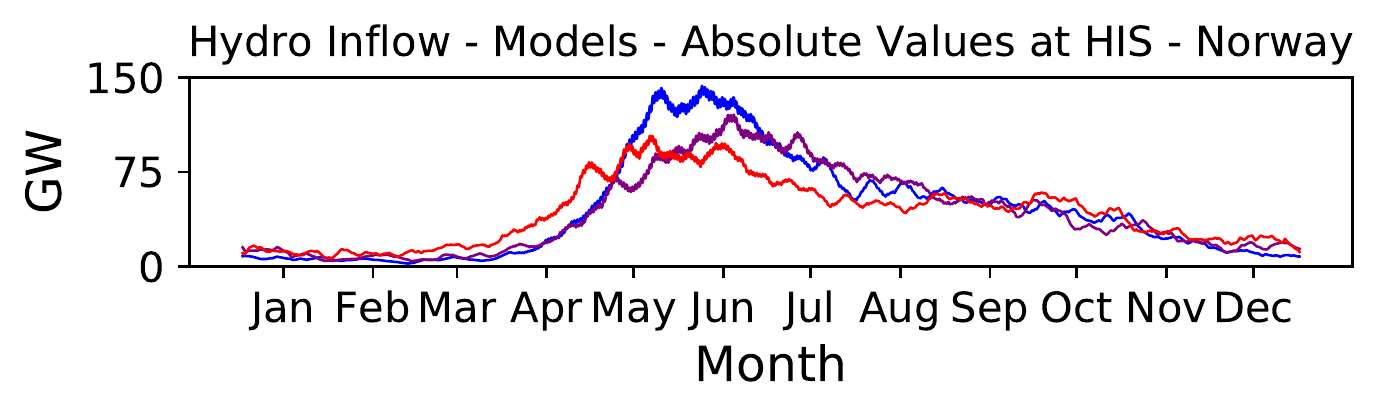}
\includegraphics[width=.49\textwidth]{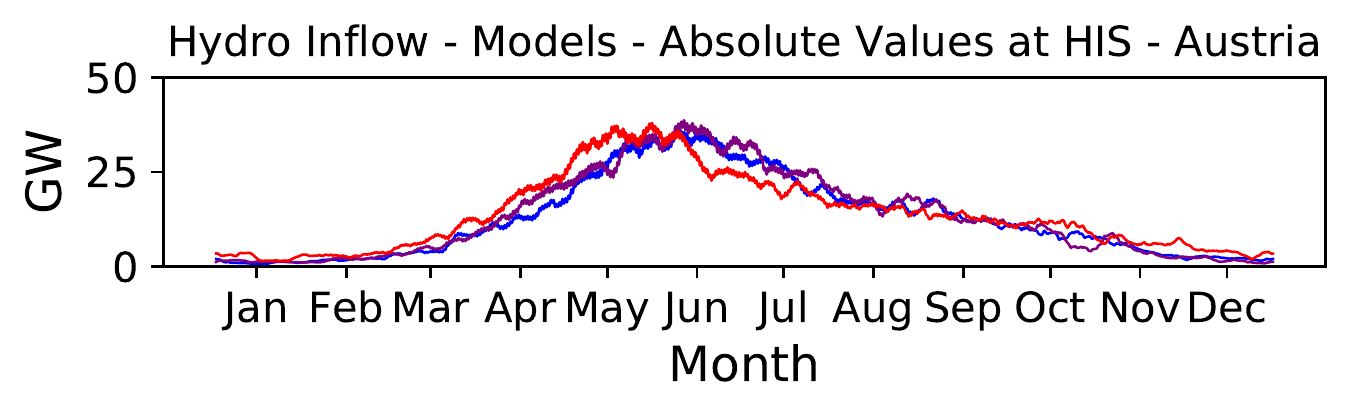}\\

\includegraphics[width=.49\textwidth]{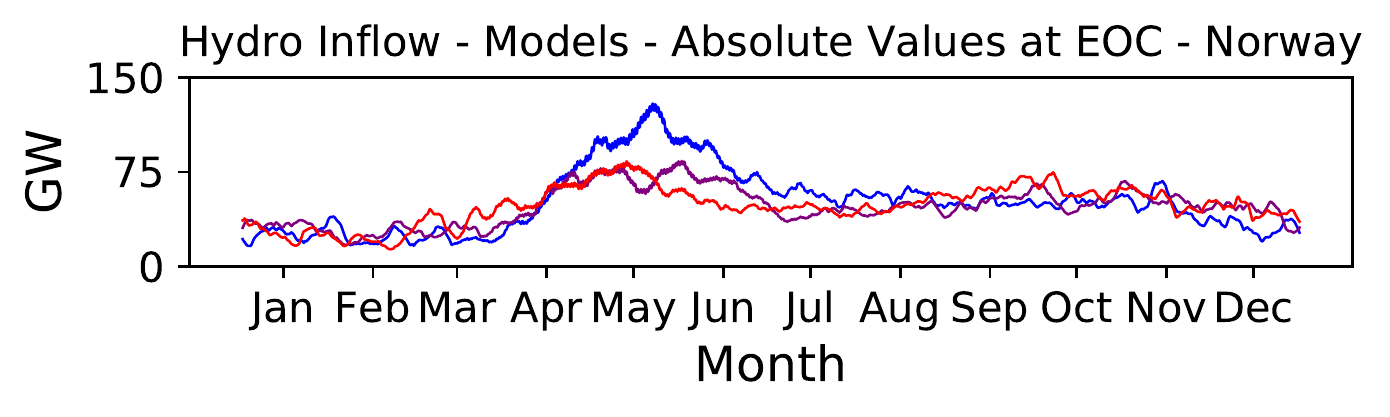}
\includegraphics[width=.49\textwidth]{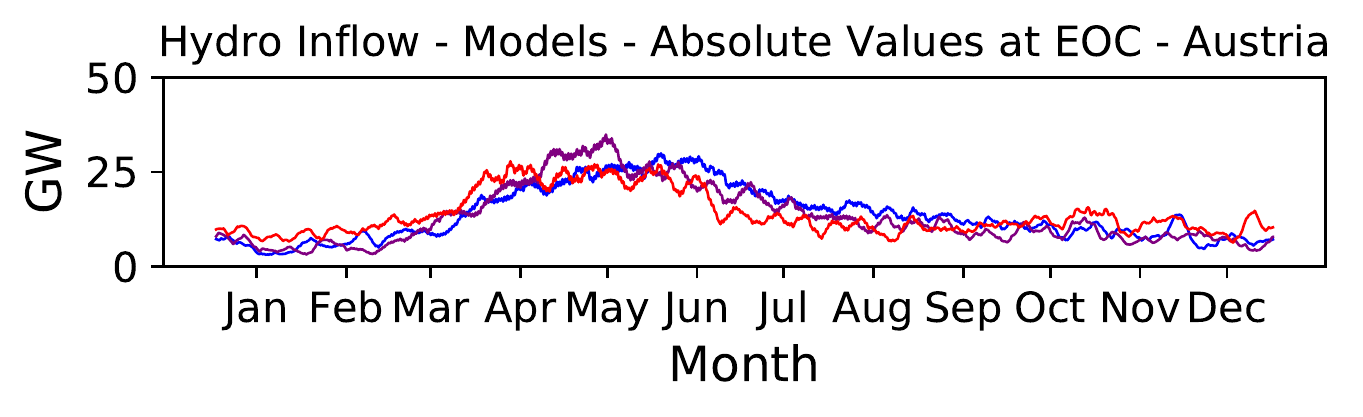}\\

\includegraphics[width=.49\textwidth]{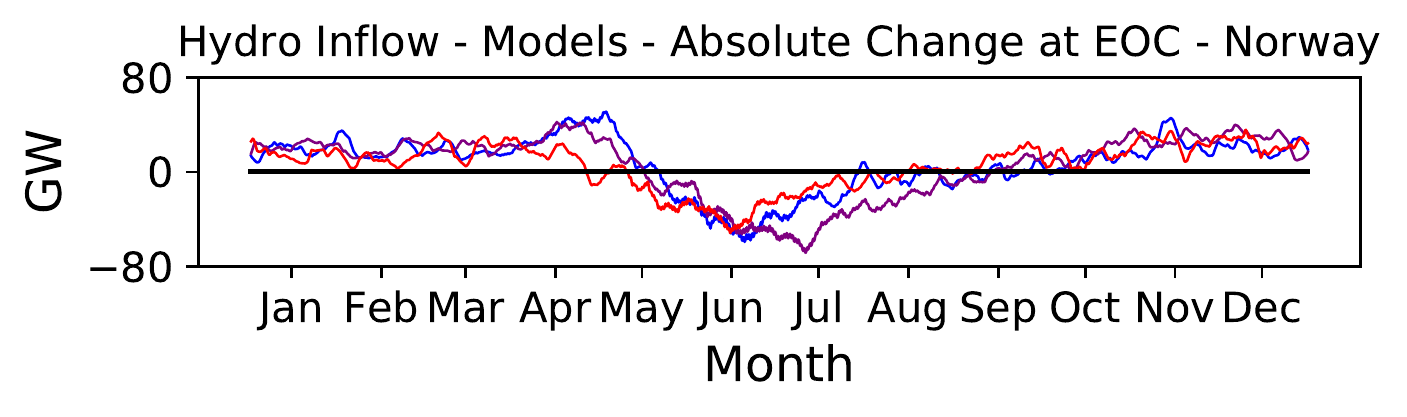}
\includegraphics[width=.49\textwidth]{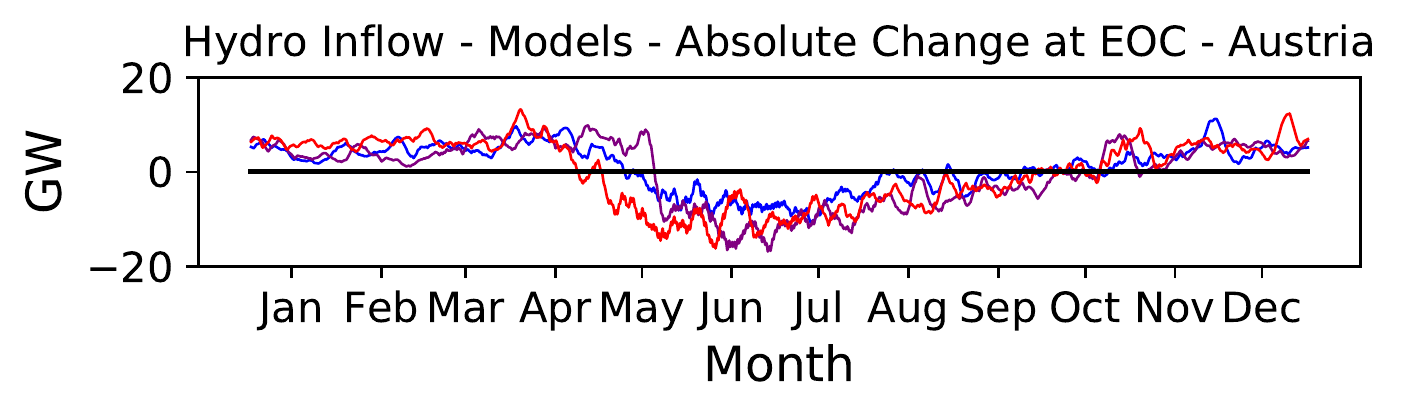}\\

\includegraphics[width=.49\textwidth]{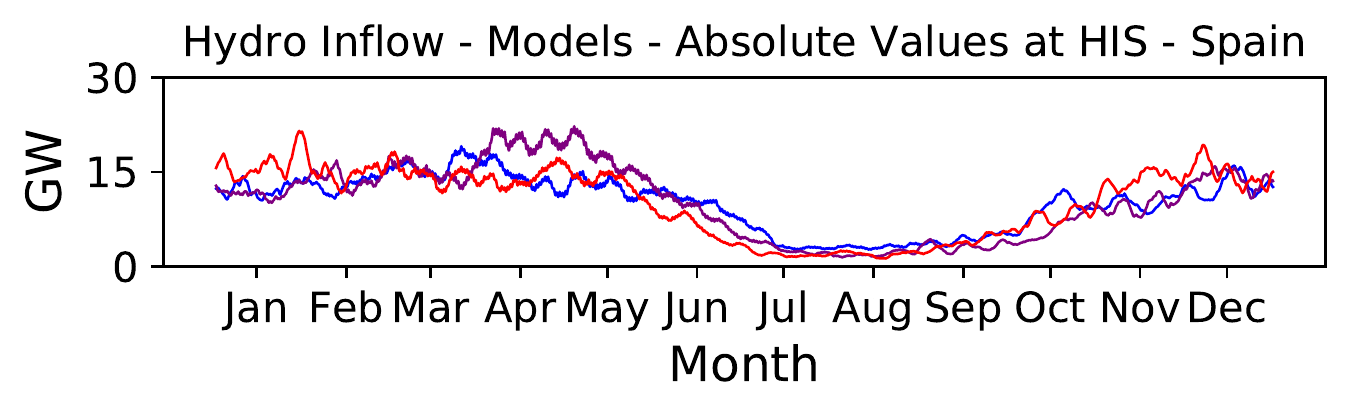}
\includegraphics[width=.49\textwidth]{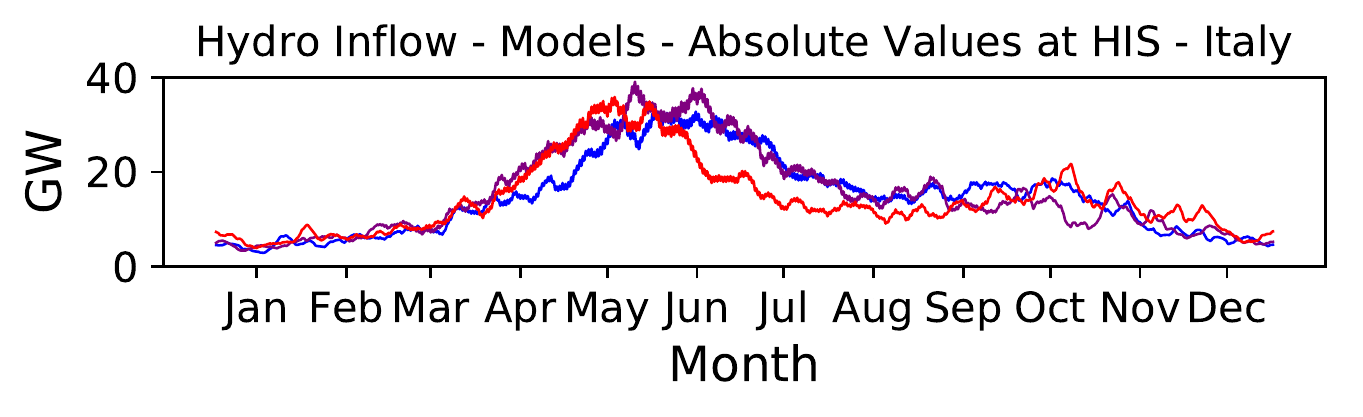}\\

\includegraphics[width=.49\textwidth]{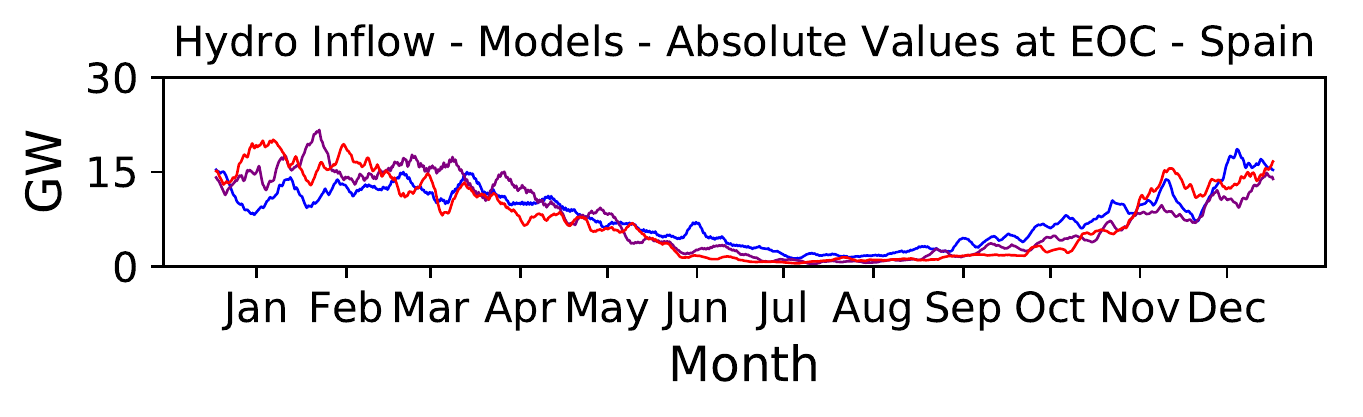}
\includegraphics[width=.49\textwidth]{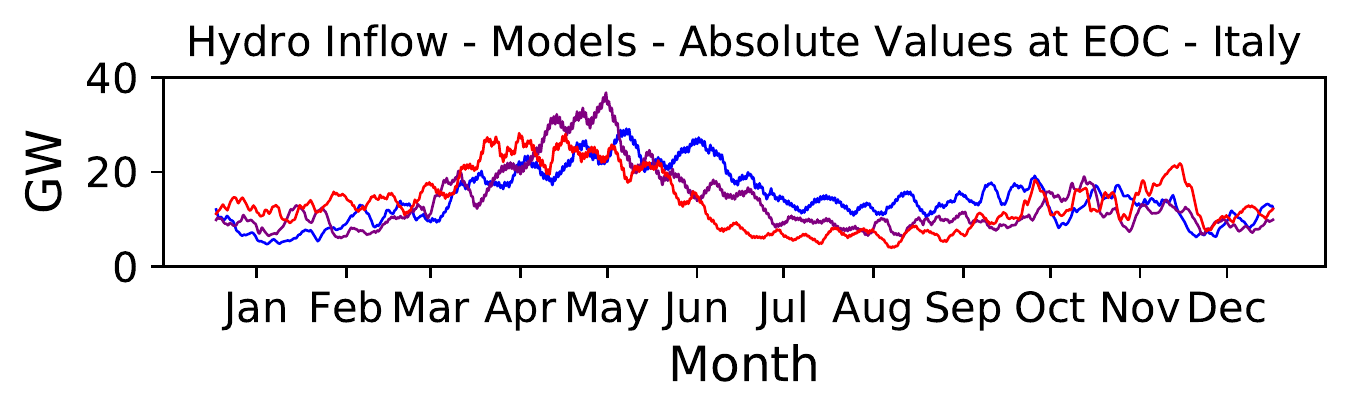}\\

\includegraphics[width=.49\textwidth]{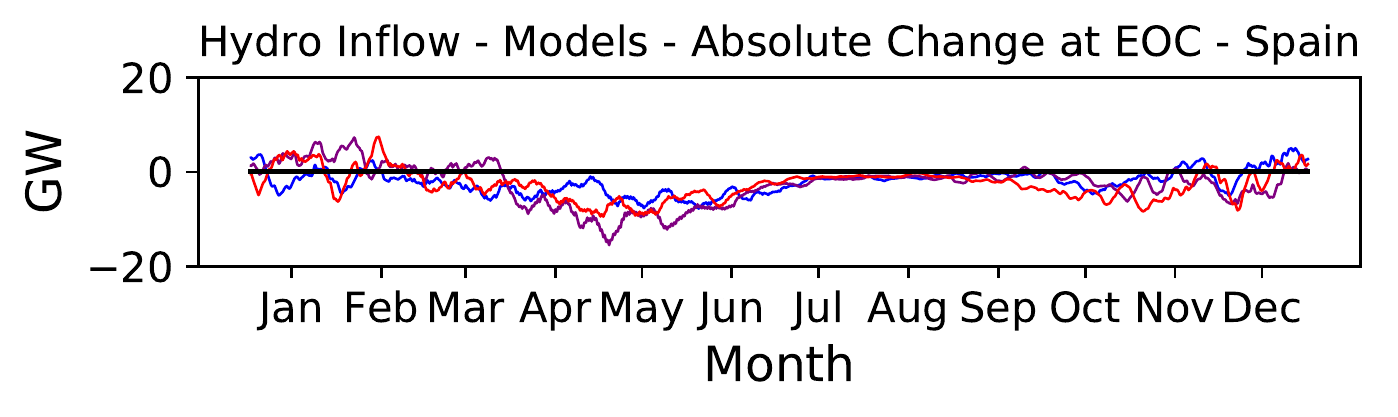}
\includegraphics[width=.49\textwidth]{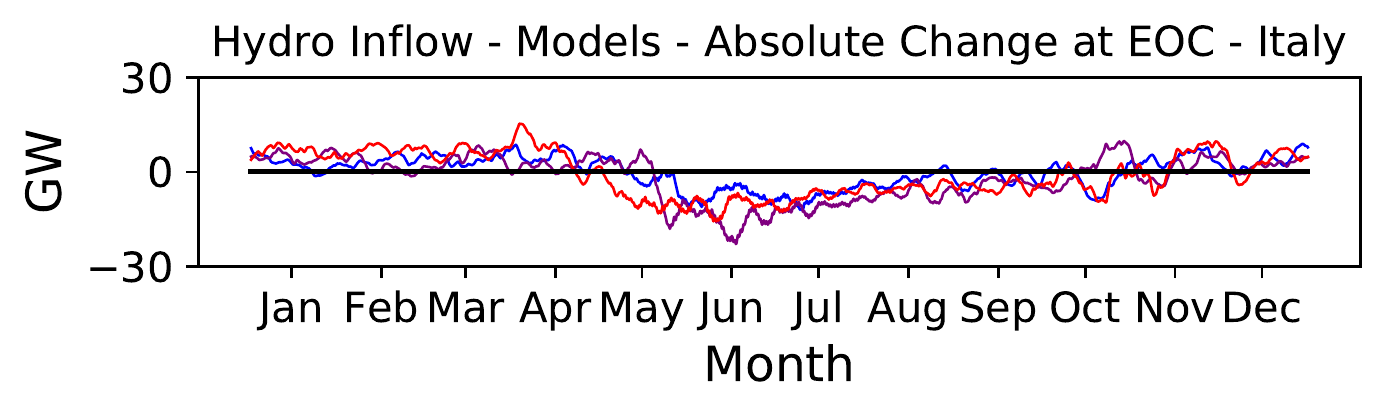}\\

\includegraphics[width=.33\textwidth]{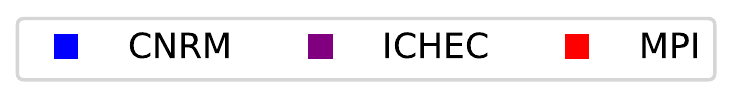}\\

\caption{\label{fig:hydro_inflow} Hydro inflow graphs for potentially available power from water run-off. Shown are the single models at the HIS period and the EOC period for the northern country Norway, the alpine country Austria, and the Mediterranean countries Spain and Italy as well as the absolute model changes at the EOC period compared to the corresponding HIS periods.}
\end{center}
\end{figure}

We focus on how seasonal hydro inflow patterns evolve throughout the century in major countries.
Seasonal patterns of hydro inflow are depicted in Fig. \ref{fig:hydro_inflow} for the four major hydro power producers Norway, Austria, Spain and Italy. The seasonal pattern of inflow in Norway, Austria, and, to some extent, Italy, shows a major peak during late spring, when snow starts to melt, and large amounts in Autumn, when rainfall is abundant. 
In the first few months of the year, inflow is almost non-existent, because temperatures are too low and snow not melting.

Changes due to climate change until the end of the century are similar for all GCMs. The spring peak is considerably reduced in size, while at the beginning and end of the year inflow increases.
This might simply be attributed to higher overall temperatures that prevent rain from freezing.

The seasonal inflow pattern of Spain looks considerably different to the other studied countries: Barely any inflow in the summer months and homogeneous amounts in the remainder of the year. Again, all GCMs show similar changes until the end of the century. In this period, differences between summer and the remainder of the year are even more pronounced than in the historical period and, in addition, overall inflow is reduced.

\section{Cost-Optimal and Highly Renewable European Power System}
\label{sec:costoptimal}

\begin{figure}[!htp]
\begin{center}
\includegraphics[width=1.0\textwidth]{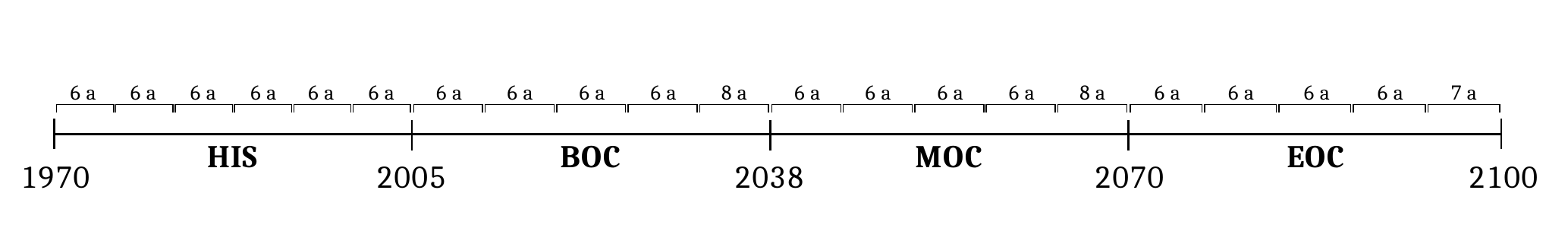}
\caption{\label{fig:optimisation_span} Intervals of the single optimisation runs.}
\end{center}
\end{figure}

To examine the impact of the change in renewable energy resources from Sec. \ref{sec:res} on a European power system, the optimisation procedure described in Sec. \ref{ss:psopt} is applied to each climate model from Table \ref{table:ensemble_members}. 
The European network actually covers a reduced set of the EU countries (as of 2017, except for Cyprus and Malta) and some non-EU countries that are associated to and take an important part in the pan-European transmission grid system: Norway, Switzerland, Bosnia and Herzegovina and the Republic of Serbia. Other peripheral countries are not considered. For a list of countries, see Table \ref{tab:network_countries}.
A single optimisation run covers an interval of 6-8 years in a three-hourly resolution as sketched in the timeline Fig. \ref{fig:optimisation_span}.


\begin{table}[p]
\begin{center}
\begin{tabular}{||c|c||c|c||}
\hline \textbf{Country Code} & \textbf{Country} & \textbf{Country Code} & \textbf{Country}\\
\hline
\hline AT & Austria & HU & Hungary\\
\hline BA & Bosnia and Herzegovina & IE & Ireland\\
\hline BE & Belgium & IT & Italy\\
\hline BG & Bulgaria & LT & Lithuania\\
\hline CH & Switzerland & LU & Luxembourg\\
\hline CZ & Czech Republic & LV & Latvia\\
\hline DE & Germany & NL & Netherlands\\
\hline DK & Denmark & NO & Norway\\
\hline EE & Estonia & PL & Poland\\
\hline ES & Spain & PT & Portugal\\
\hline FI & Finland & RO & Romania\\
\hline FR & France & RS & Republic of Serbia\\
\hline GB & United Kingdom & SE & Sweden\\
\hline GR & Greece & SI & Slovenia\\
\hline HR & Croatia & SK & Slovakia\\
\hline
\end{tabular}
\caption{Overview of all countries taking part in the European network - each country is represented by one node} \label{tab:network_countries}
\end{center}
\end{table}

\subsection{Installed Capacities}
\begin{figure}[!htp]
\begin{center}
\includegraphics[width=.42\textwidth]{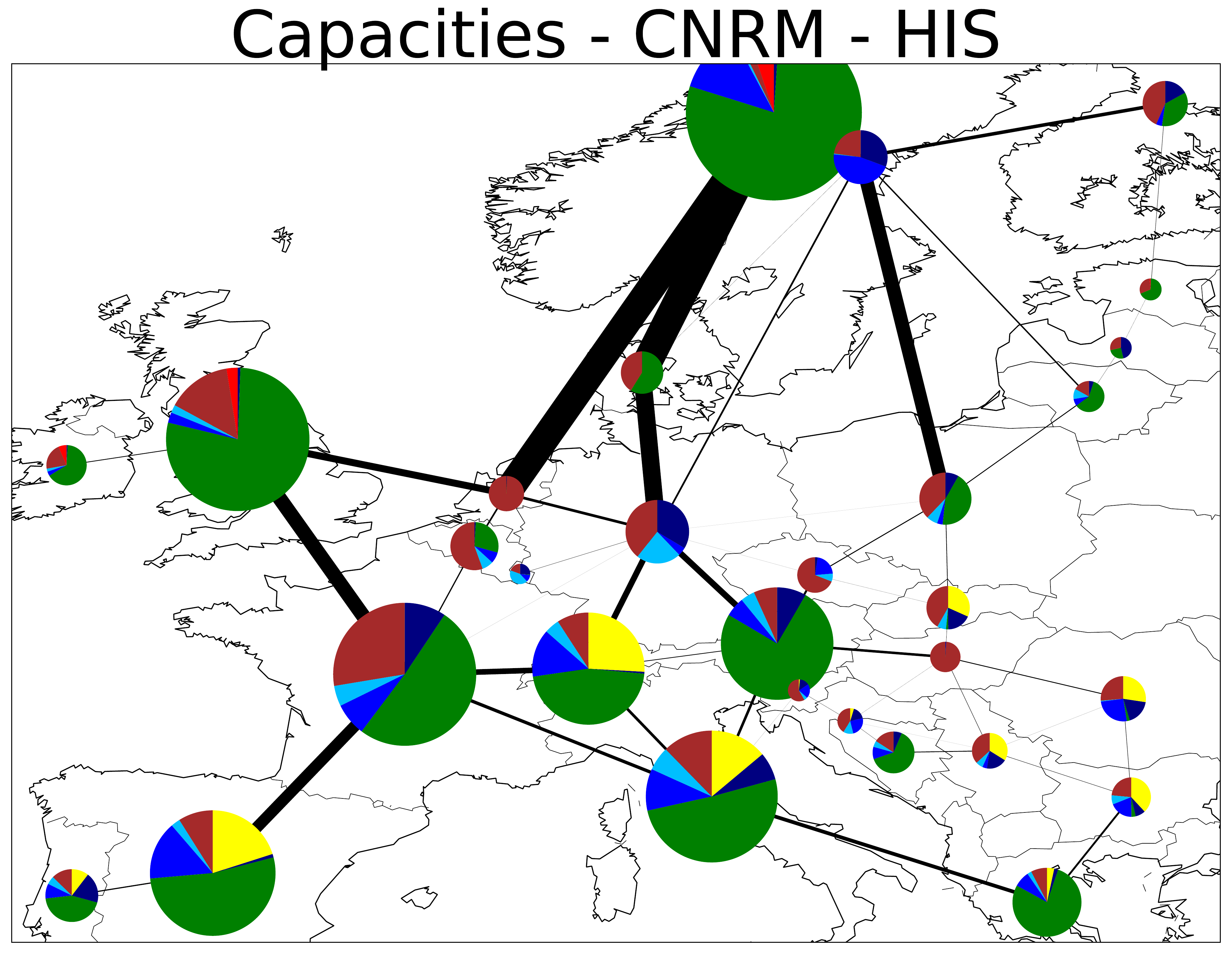} \quad
\includegraphics[width=.42\textwidth]{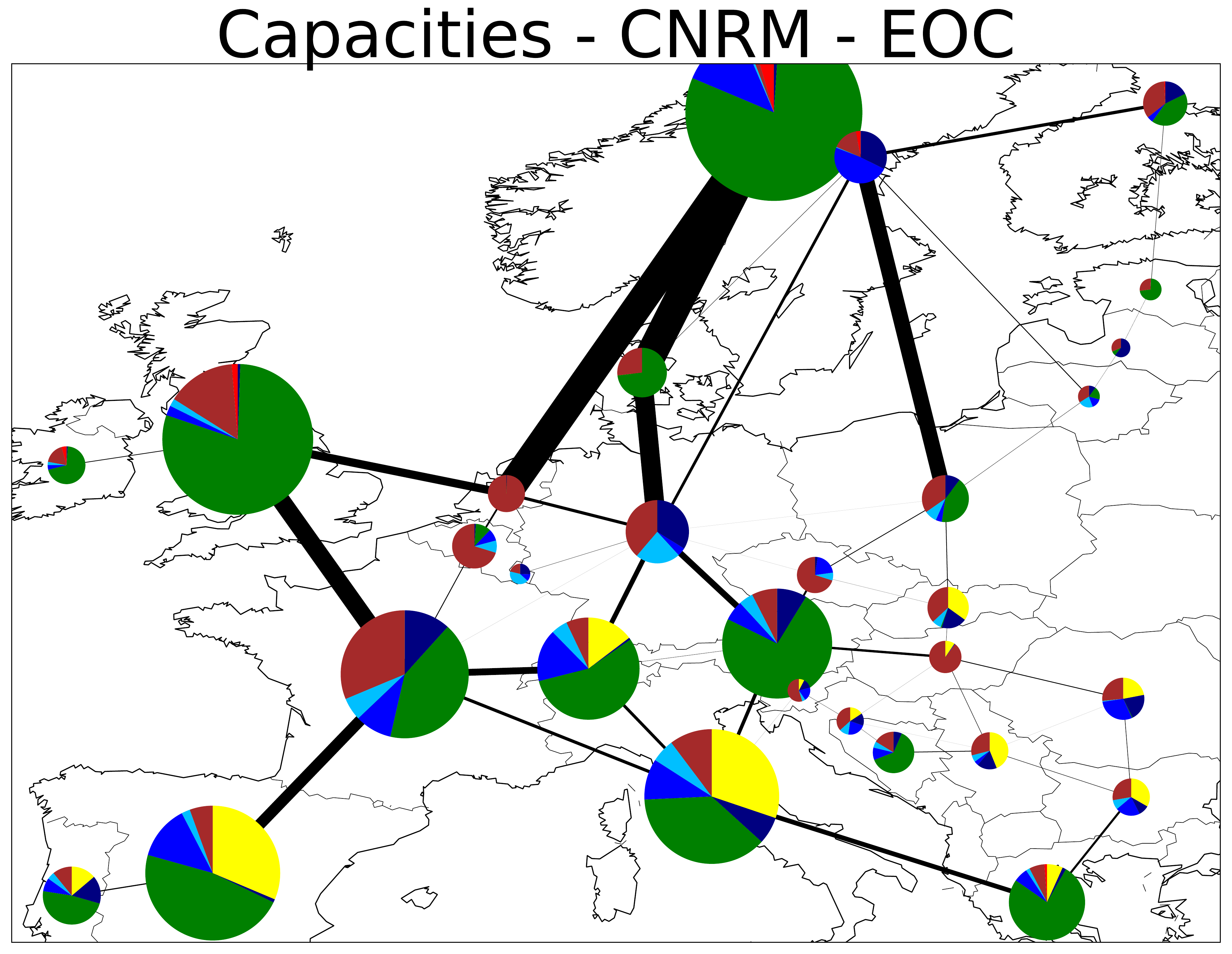}\\

\includegraphics[width=.42\textwidth]{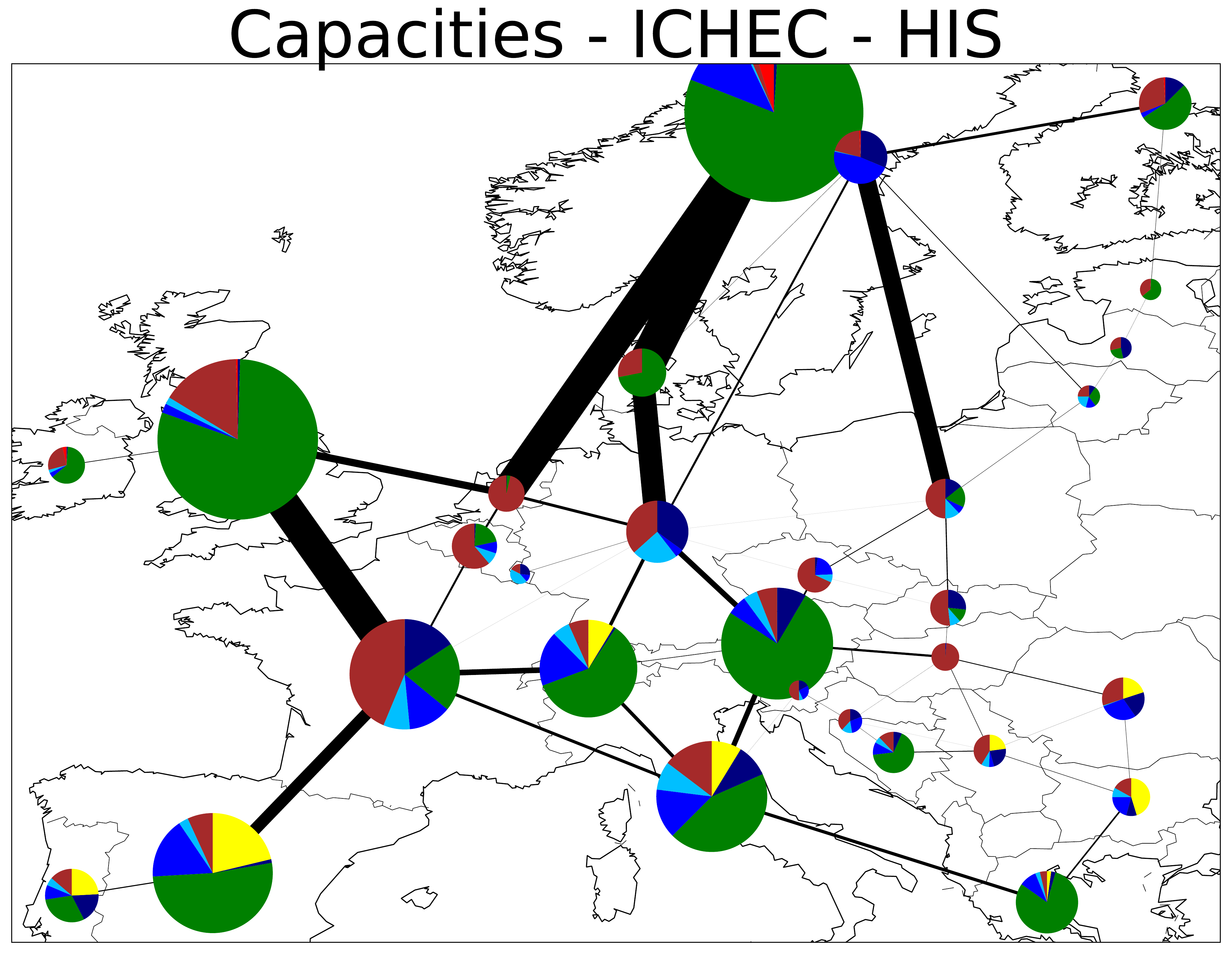} \quad
\includegraphics[width=.42\textwidth]{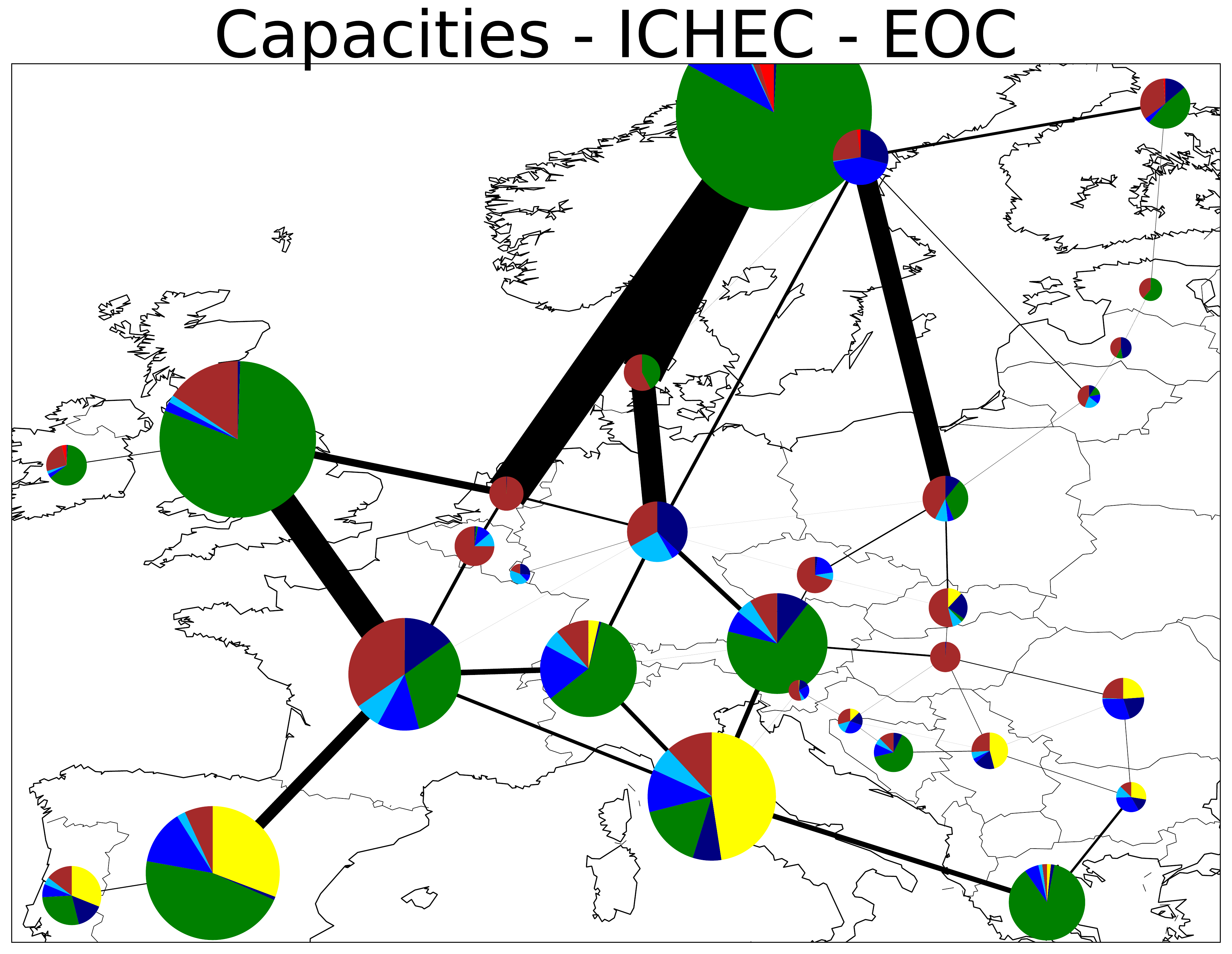}\\

\includegraphics[width=.42\textwidth]{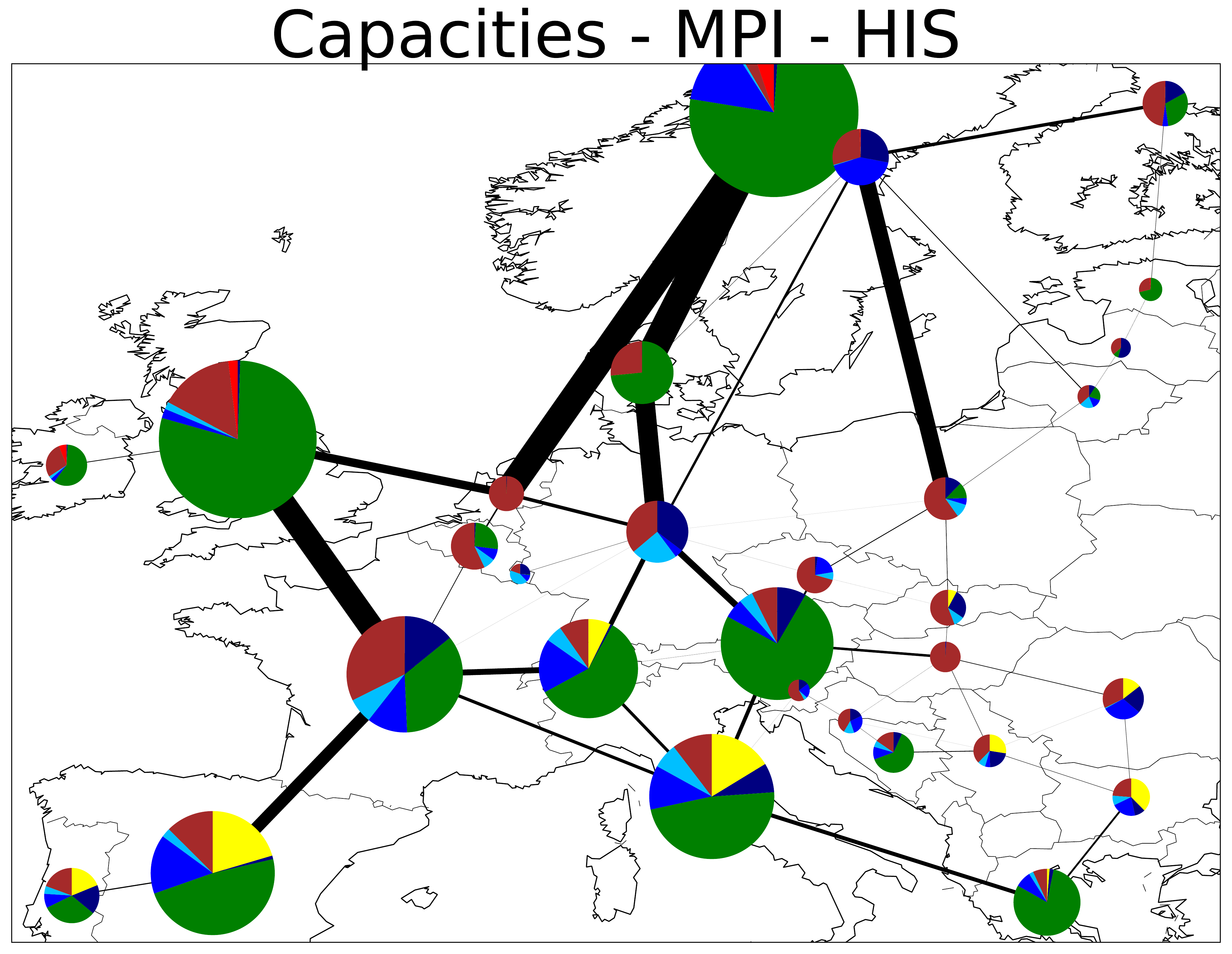} \quad
\includegraphics[width=.42\textwidth]{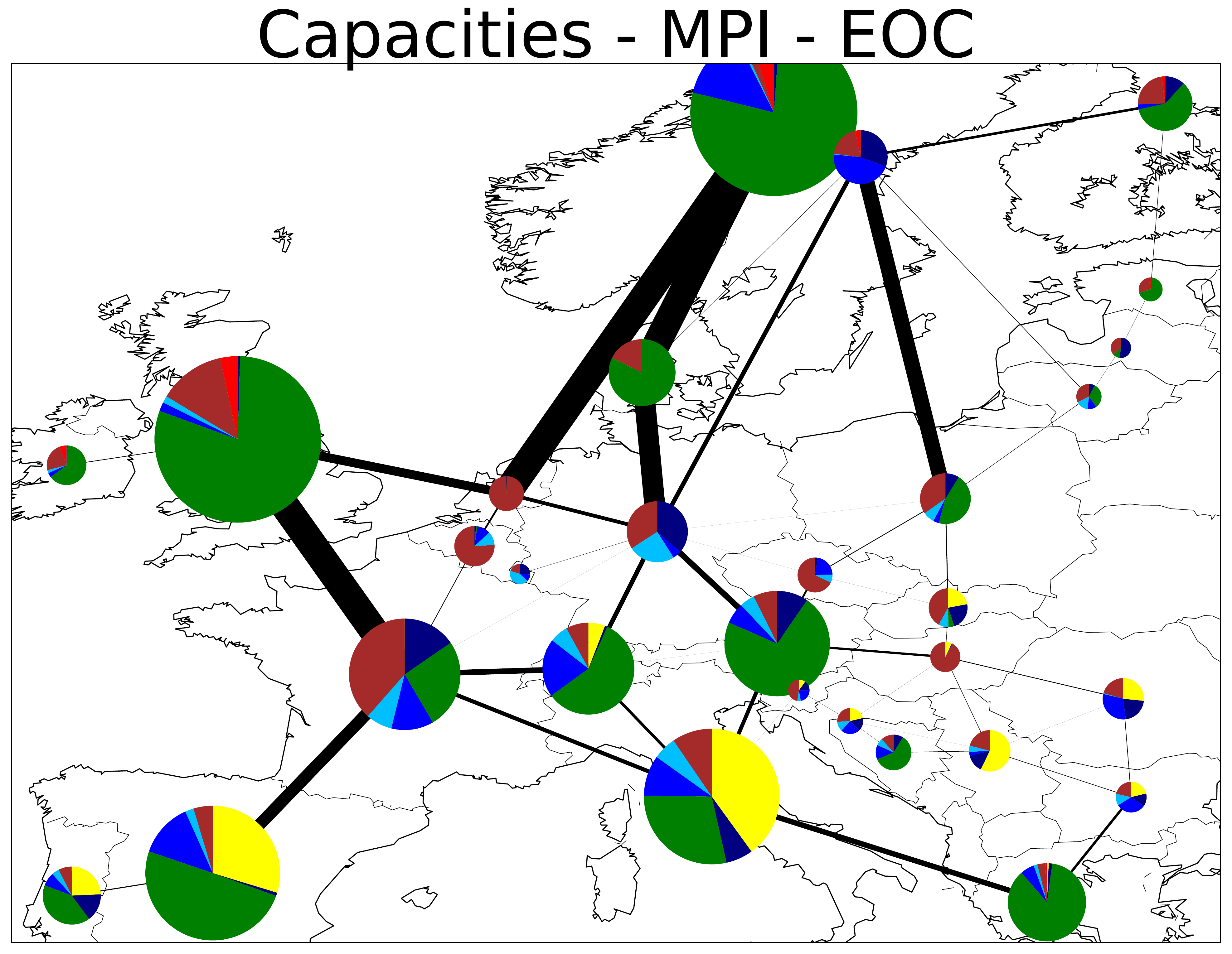}\\

\includegraphics[width=.9\textwidth]{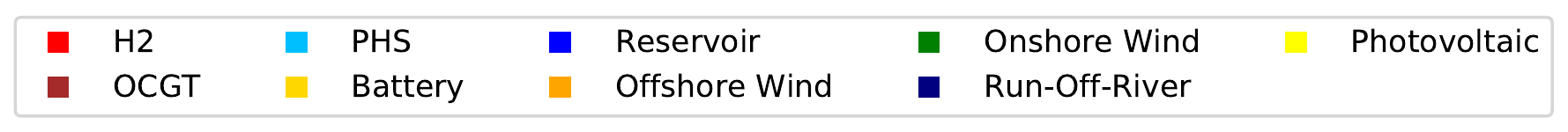}
\caption{\label{fig:cap_dist_pie} Installed capacities in the cost-optimal case for all energy carriers, storage technologies and lines; in the first column single models at the HIS period and in the second column single models at the EOC period are shown.}
\end{center}
\end{figure}

Fig. \ref{fig:cap_dist_pie} shows the optimised installed capacities of generators, storage units and transmission lines in Europe for the HIS period and the EOC period resulting from all three ensemble models. Hydro power plant capacities are not expandable in the model and set to today's values; it is assumed that geospatial potential for dams and run-off-river plants in Europe are already well exploited. Therefore, one cannot simply quantify the change in installed capacities as for the carrier types wind and solar. Instead, actual averaged generation due to changes in surface runoff is analysed in the following. \\
 

\underline{Historical Epoch (HIS)}\\
In the historical epoch, it is evident that onshore wind is the most important power source and mainly deployed in  Norway, Great Britain, France and the Alpine region, but also in Spain and Italy. This is in line with the observed high capacity factors for onshore wind in those regions.
The overall picture is very similar for all models.
On the other hand, offshore wind capacities are barely represented.  

PV modules are installed only in southern and south-eastern Europe, where capacity factors are high, thus rendering their installations cost-effective.

OCGT capacities are built all across Europe in small amounts as backup and smoothing facilities. The global CO$_2$ constraint puts a limit on total dispatch from gas turbines and thus on capacities.

Hydro capacities are set to today's capacity values and assumed to be well-exploited and non-expandable. The effect of climate change on energy generation from hydro sources is discussed in the next subsection.

Storage units are only deployed in very small amounts. Major shares fall upon PHS in countries suitable for this storage type, i.e., France, Italy and Germany.

The cost-optimal solution determines an expansion of the entire inter-connecting transmission line capacity system by a factor of 9.35 compared to today's aggregated value (derived from NTC) of 30 TWkm. 
However, one should note that 
i) this value does not equal the existing thermal limit of inter-connecting transmission capacities and
ii) the need for additional capacities is over-estimated by a model, where countries are aggregated to single nodes \cite{horschpypsa}. If the resolution is increased, the need for additional transmission capacities shrinks. \\
Transmission line capacities concentrate in northern Europe, which reflects the fact that correlation of wind feed-in decreases over distances of hundreds of kilometers and therefore wind power benefits from a well expanded transmission grid. Contrary to this, transmission capacities in southern and southeastern Europe are comparably weak. Here, PV feed-in is favoured due to higher capacity factors, but corresponding generation is mainly determined by the deterministic diurnal cycle and therefore strongly correlated.
Optimisation scenarios with lower transmission caps demonstrate that lowering the global transmission line limit increases the deployment of storage, especially batteries, as shown by Schlachtberger et al. \cite{schlachtberger2017benefits}.

To summarise, Europe can be roughly divided into two parts. The north and north-west, which are dominated by onshore wind and a strongly reinforced transmission grid; and the south and south-east with a more diversified capacity distribution, favouring larger PV shares and a comparably weak transmission grid. As will be seen in the following subsection, climate change does not alter this pattern, but increases the importance of PV in the south and south-east significantly.\\

\underline{End of Century (EOC)}\\
At the end of the century, two major climate change effects are visible:

First, onshore wind distribution and the associated transmission grid expansion in the north and north-west of Europe remains stable, showing only small capacity decreases around the Netherlands and Belgium as well as the Baltic countries and the Alpine region. The main onshore wind production centers Norway and Great Britain, on the other hand, show even small increases in deployed onshore wind capacities. Despite most countries suffer small decreases in onshore wind capacity factors, the carrier type remains essential all across Europe. Onshore wind installations even grow from a total of 664.05 GW at the HIS period by 0.8\% up to a value of 669.42 GW at the EOC period and stay nearly constant.

Second, installed PV capacities in the south and south-west gain huge increases and are expanded in almost every associated country, especially in Spain and Italy, but also in the Balkans, while the transmission grid in those regions still remains weak. This is in accordance to the changes in PV capacity factors. PV installations grow from a total of 65.57 GW at the HIS period by 82\% up to a value of 119.42 GW at the EOC period.

Total changes in deployed offshore wind capacities are negligibly small and, as in the HIS period, almost no offshore wind is deployed in the EOC period despite some observed positive changes in capacity factors. Again, offshore wind seems not to be cost-competitive under the given assumptions.

The overall small deployment of OCGT facilities is only slightly variable, mostly dependent on the cost-optimal distribution of backup and smoothing abilities in the given framework, but less on climate change impact.

Due to its largely uncorrelated feed-in on the synoptic scale, i.e. due to the wind speed's small correlation lengths compared to the network size, the integration of wind generation is stronger supported by long-distance transmission capacities than by storage units. However, in Norway and Great Britain also storage types suitable for seasonal variance patterns, i.e. hydrogen storage, are deployed. Hydrogen storage installations grow from a total of 14.93 GW at the HIS period by 7\% up to a value of 15.93 GW at the EOC period, whereas battery storage only gets deployed in the MW range, independent of the HIS period or the EOC period.

The growing shares of PV in areas with a weak transmission grid such as south and south-east Europe are likely justified by the growths of the associated capacity factors, which favours the deployment of PV capacities. Since its power generation's variances reflect the deterministic and naturally strongly correlated diurnal patterns of its source, i.e. the incoming irradiation, this trend also profits from increased installations of storage units.

Finally, the transmission grid shows some, but overall small response to climate change. Until the EOC period its total capacity increases by about 6\% compared to the HIS period.

Studying the transmission line capacities of single optimisation runs within the time spans indicated in Fig. \ref{fig:optimisation_span}, one observes considerable timespan-to-timespan fluctuations of the optimal transmission capacities up to 100 TWkm, independent of the chosen model. Although the general trend of rising transmission needs under climate change stays apparent, this might indicate a rather flat cost optimum with respect to the transmission variable.\\

\underline{Model Differences}\\
Overall, the differences in deployed generation, storage and transmission capacities and  among the different climate models are rather small.
On country scale, onshore wind installations show small model dependent variations with changes in both directions, while the trend for offshore wind is, due to the very small overall deployment, mostly negligible.
PV installations on the other hand show a strong positive trend for the MPI and CNRM model especially in Southern Europe, but for the ICHEC model increases are less pronounced and in some northern countries even small decreases are observed.
OCGT facilities show only small changes in both sign directions across all models, which can be understood by the role of backup and smoothing renewables as well as by the restricted generation under a global emission constraint.

\begin{figure}[!htp]
\begin{center}
\includegraphics[width=.4\textwidth]{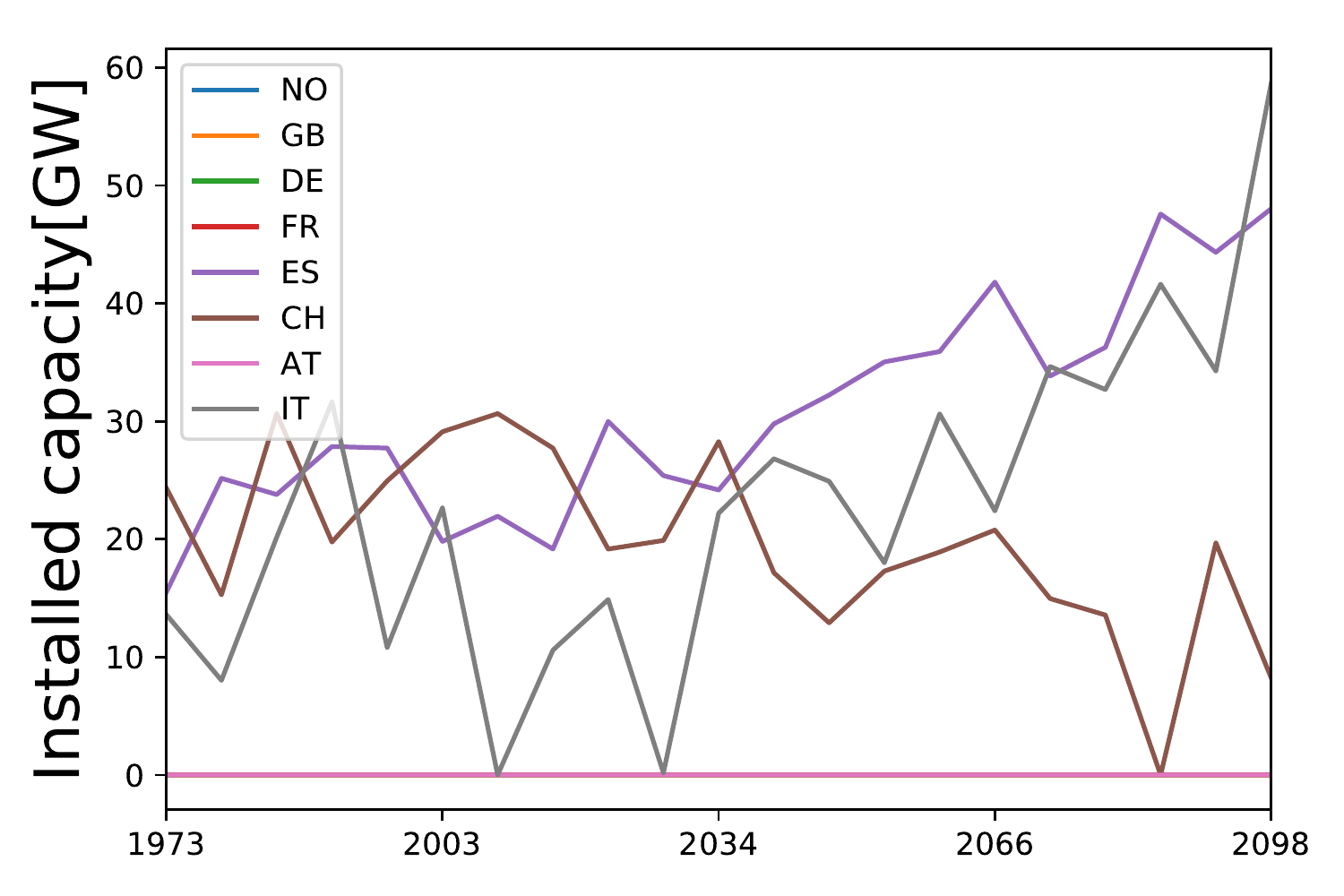}\includegraphics[width=.4\textwidth]{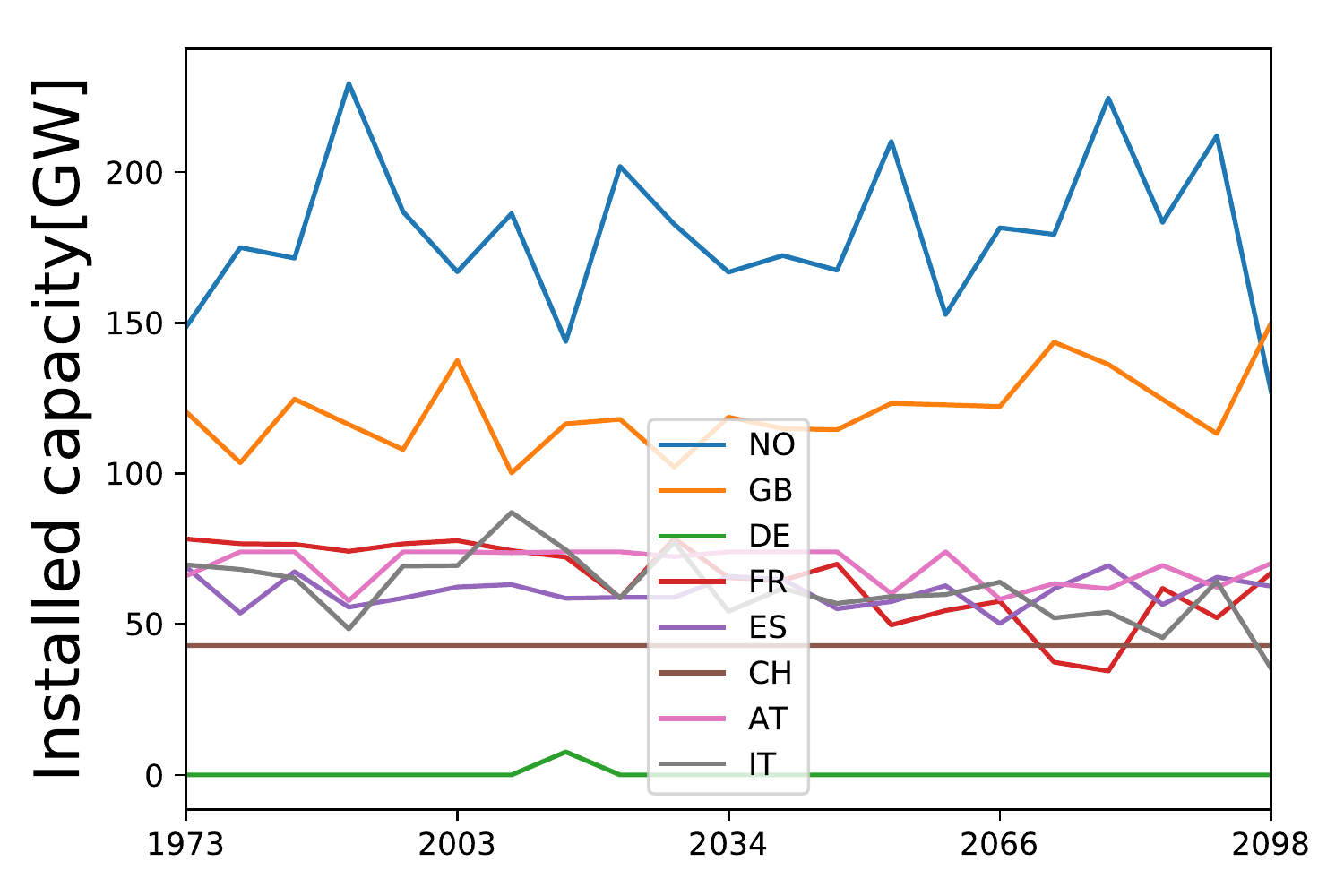}\\
\includegraphics[width=.4\textwidth]{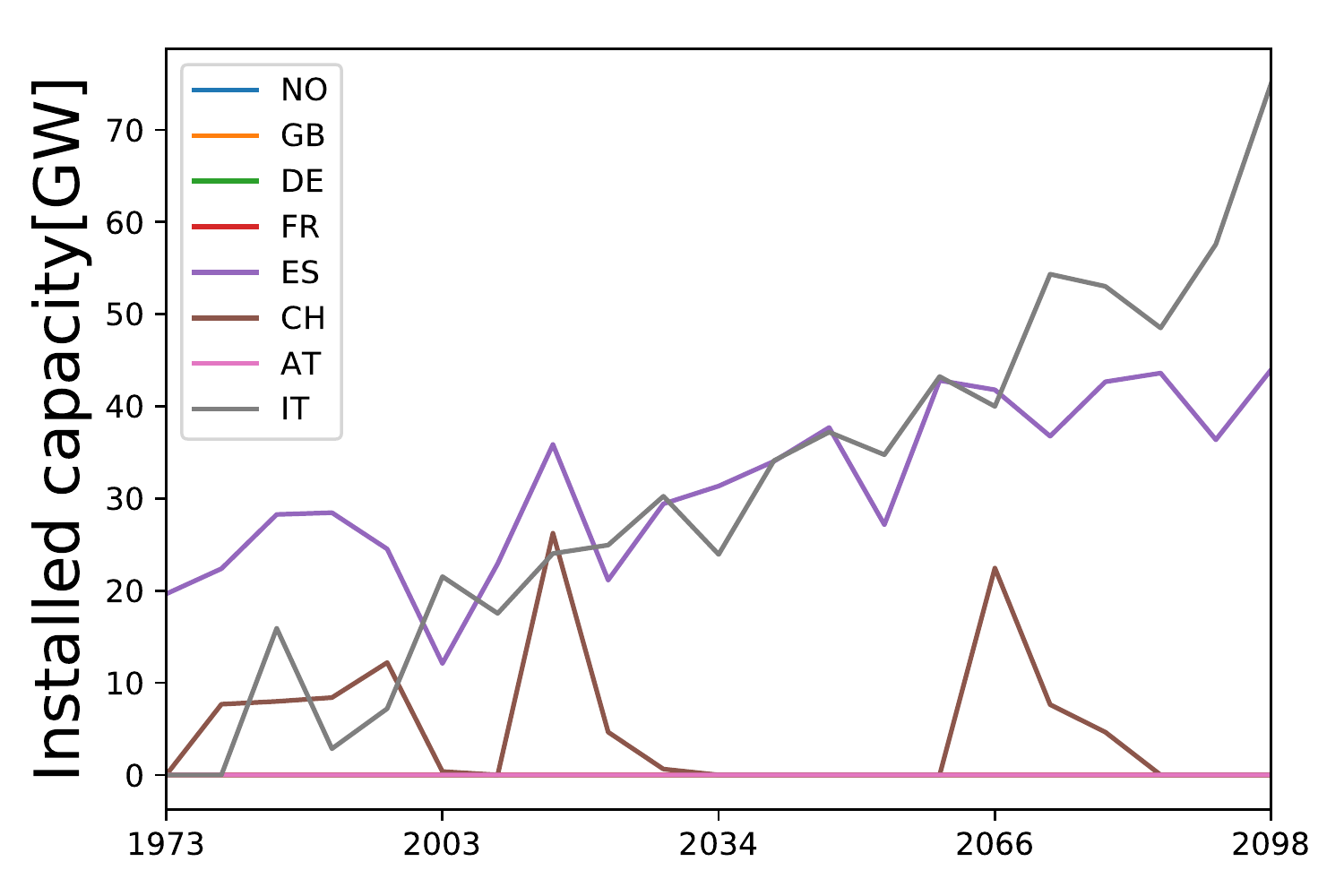}\includegraphics[width=.4\textwidth]{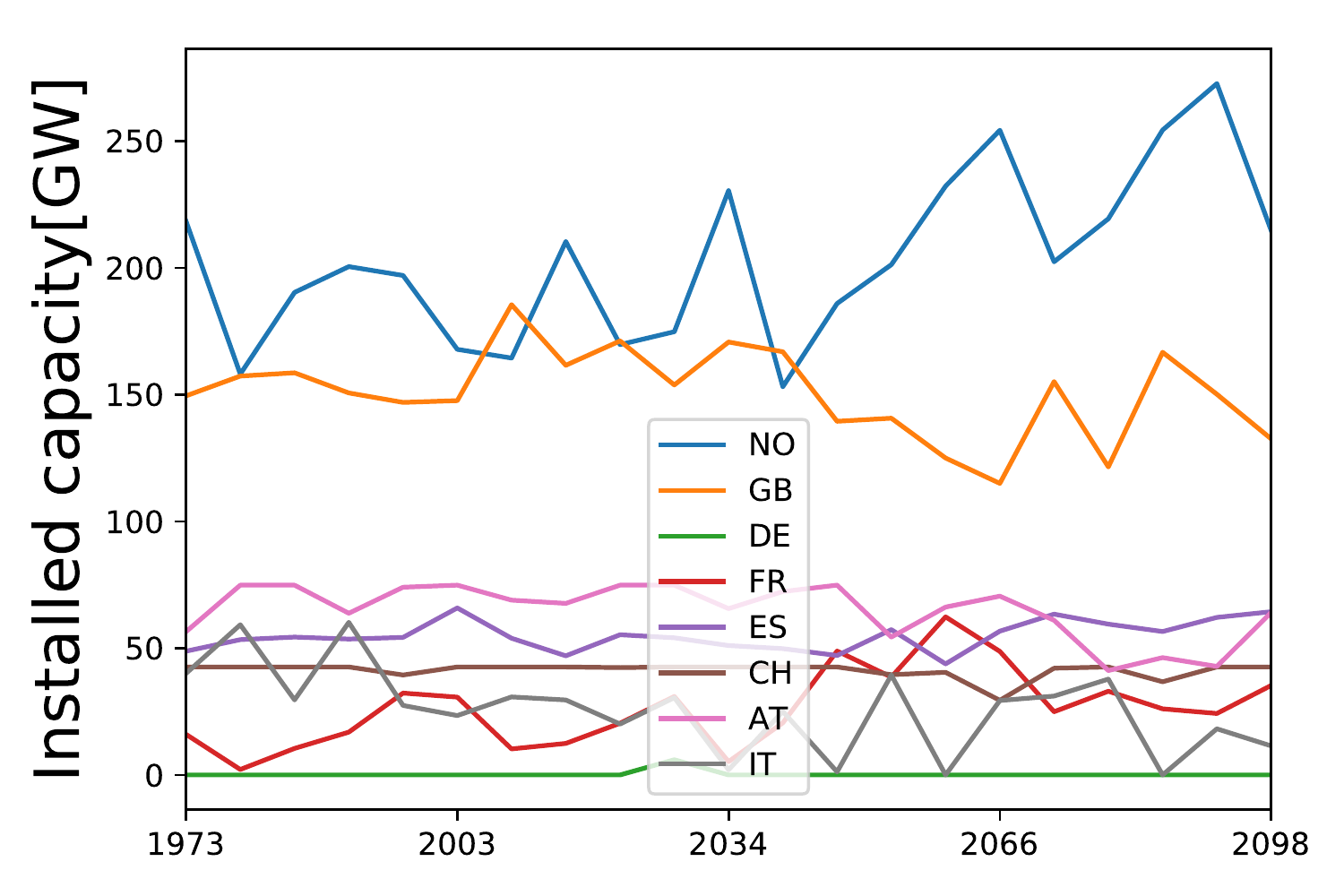}\\
\includegraphics[width=.4\textwidth]{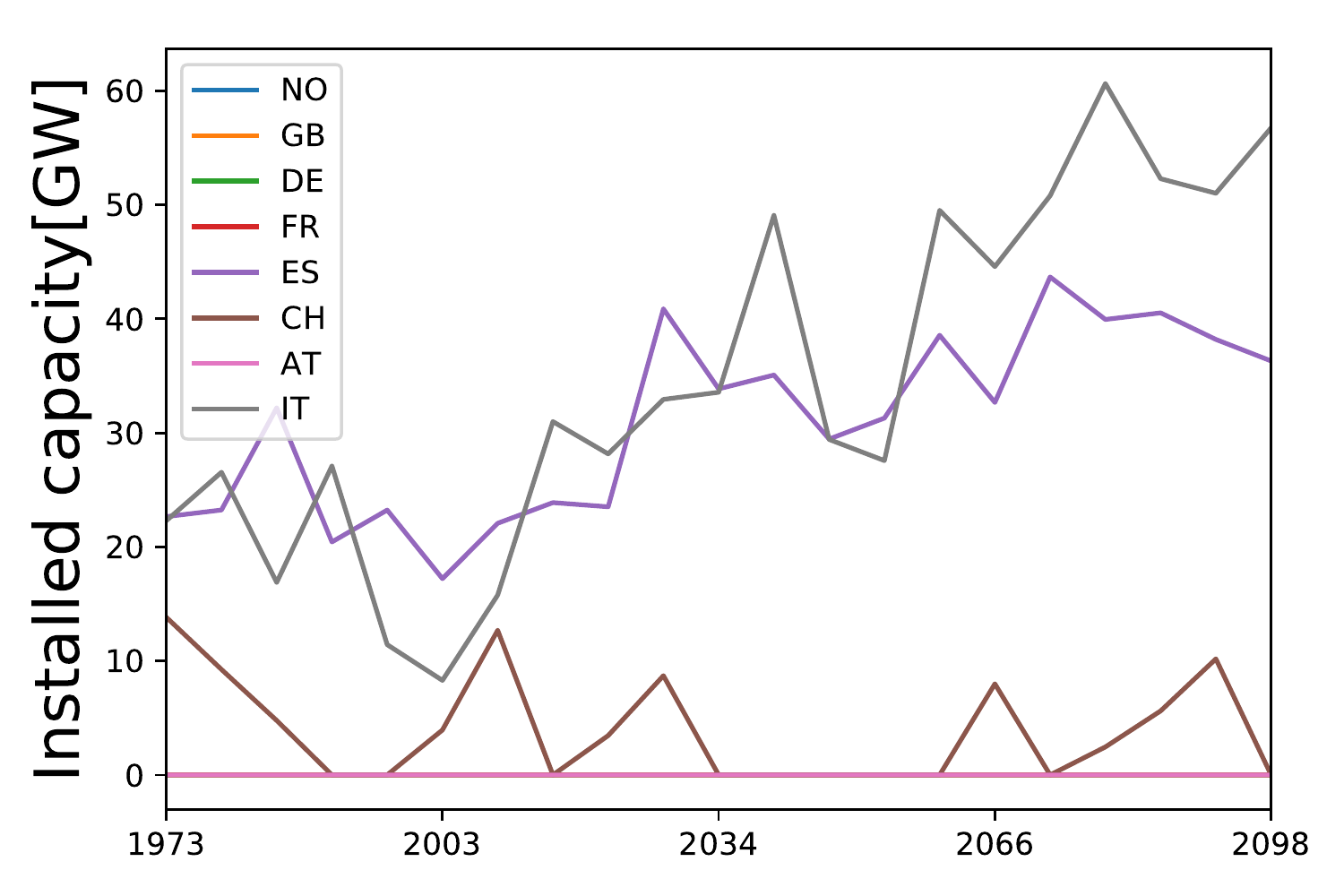}\includegraphics[width=.4\textwidth]{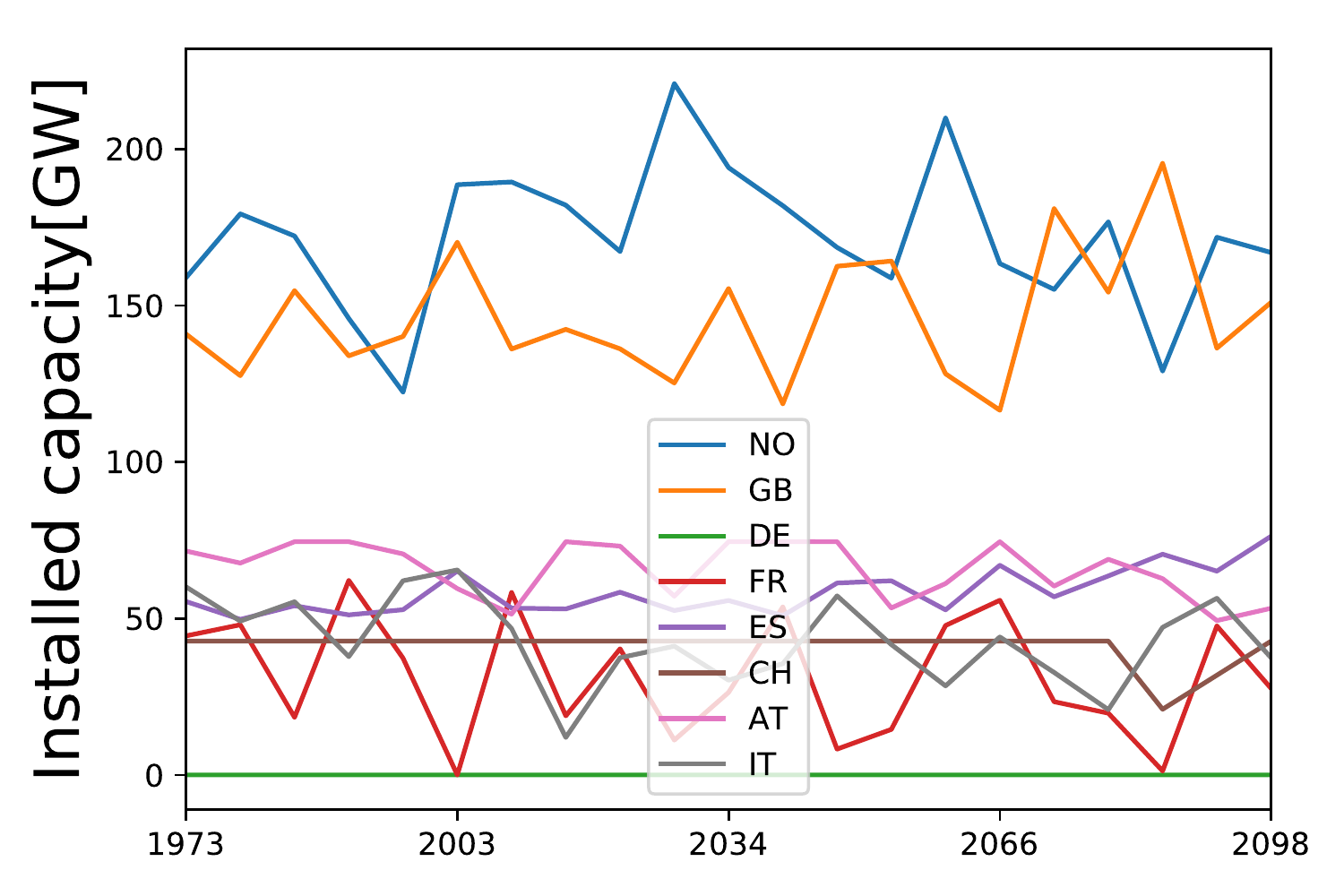}
\caption{\label{fig:cap_development} Installed capacities as a function of time for solar PV (left) and onshore wind (right) for several selected countries.
From top to bottom: CNRM, ICHEC, MPI.}
\end{center}
\end{figure}

Fig. \ref{fig:cap_development} shows installed capacity as a function of time for solar and onshore wind for all three models in eight chosen countries.
In general, models behave similarly: For wind, all countries show fluctuations, but no clear trends are apparent.
However, for solar PV Italy and Spain show a clear trend of increasing PV capacities in all three models, while strong fluctuations for Switzerland are observed.
For each period, the problem is solved as an green field investment problem, i.e., without initial capacities.
The fluctuations that can be observed, for instance for PV capacities in Switzerland, are likely caused by the difference in capacity factors of PV and wind between periods. 
´

\subsection{Hydro Generation} \label{sec:hydro_generation}
\begin{figure}
\begin{center}
\includegraphics[width=.9\textwidth]{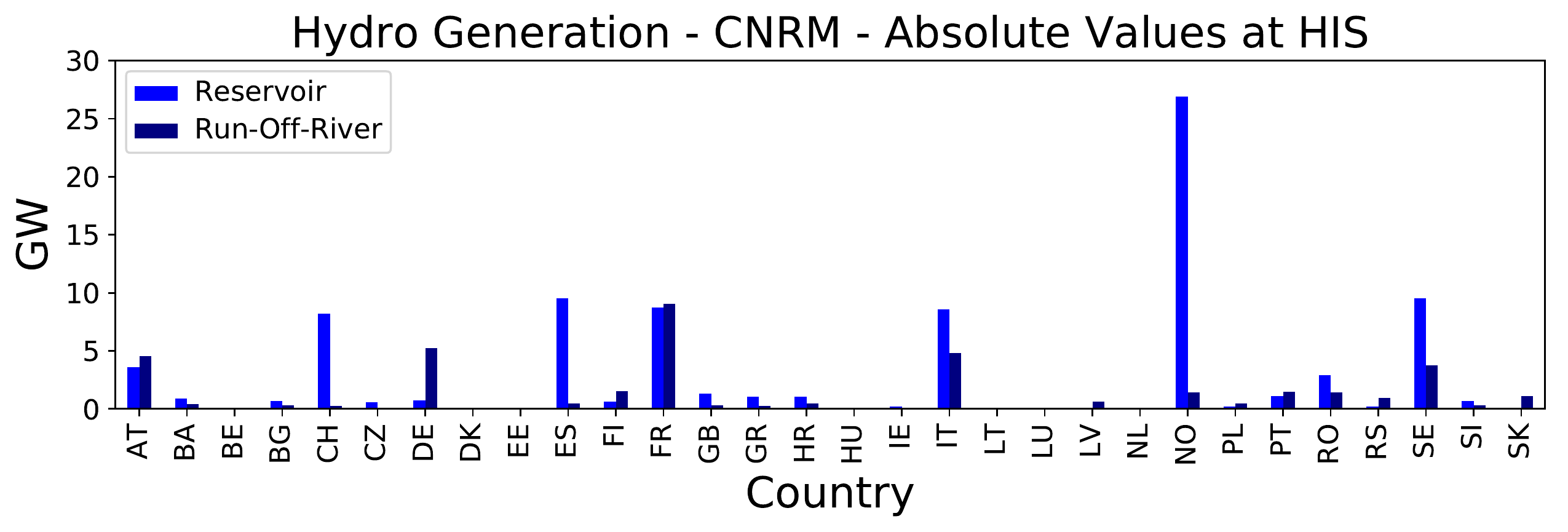}
\includegraphics[width=.9\textwidth]{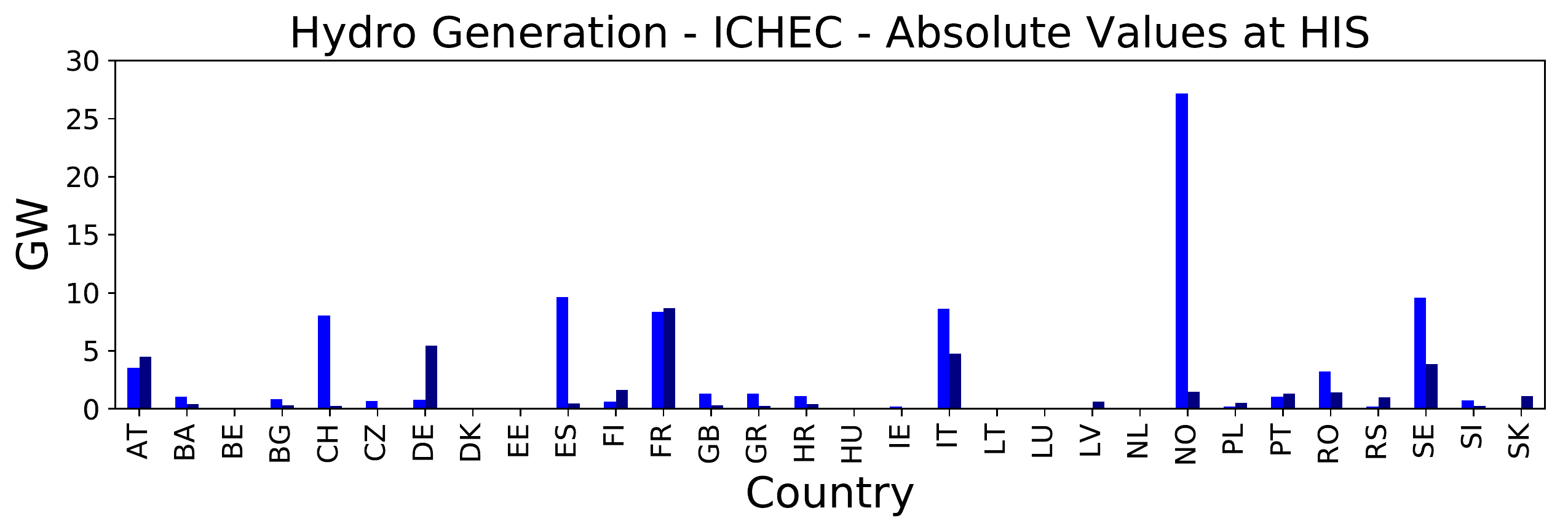}
\includegraphics[width=.9\textwidth]{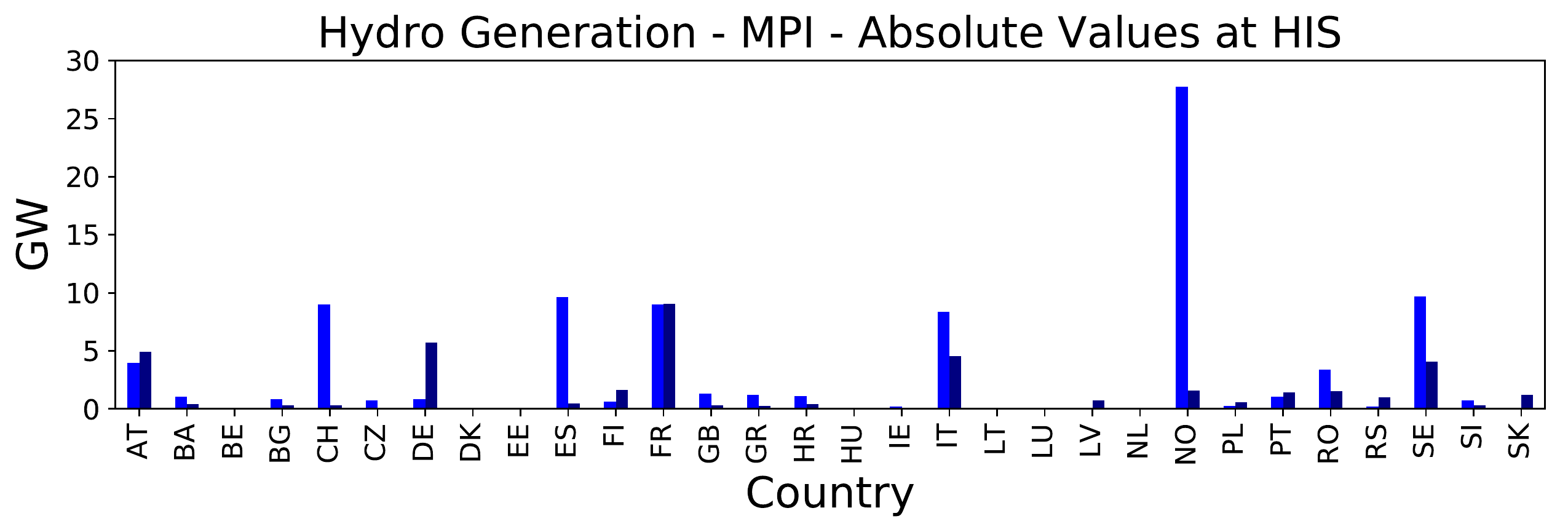}
\caption{\label{fig:hydro_generation_absolute} Hydro power generation in the cost-optimal case for reservoir (blue) and run-off-river (navy) plants; shown are the single models at the HIS period.}
\end{center}
\end{figure}

\begin{figure}
\begin{center}
\includegraphics[width=.9\textwidth]{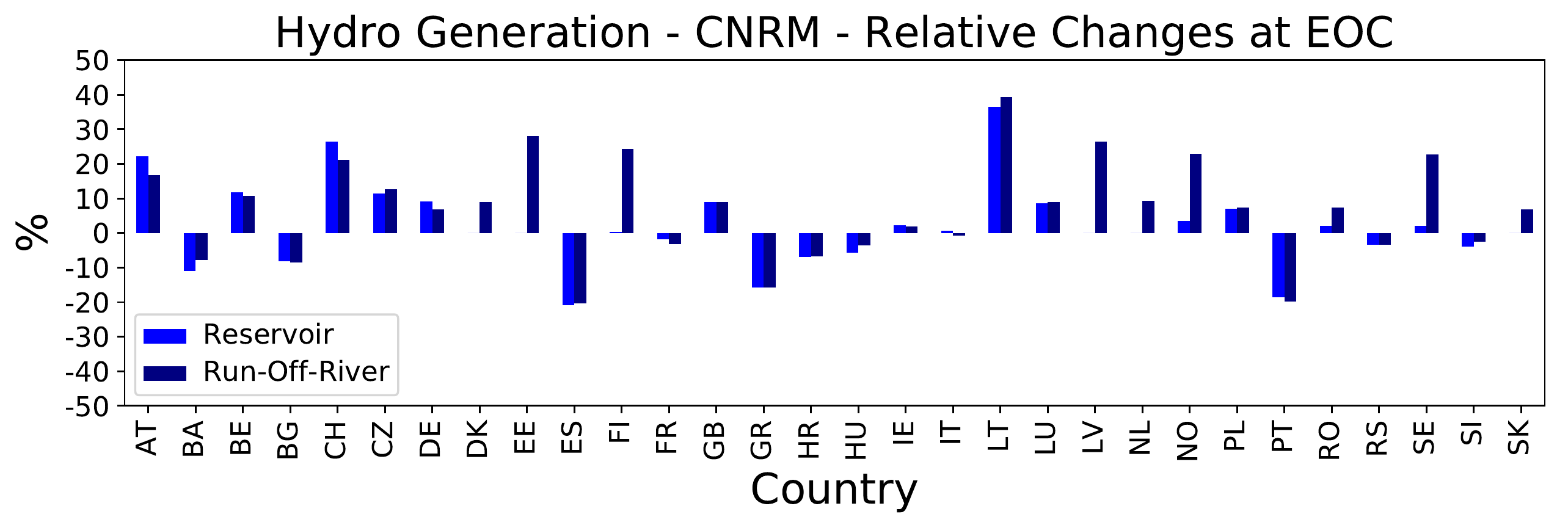}
\includegraphics[width=.9\textwidth]{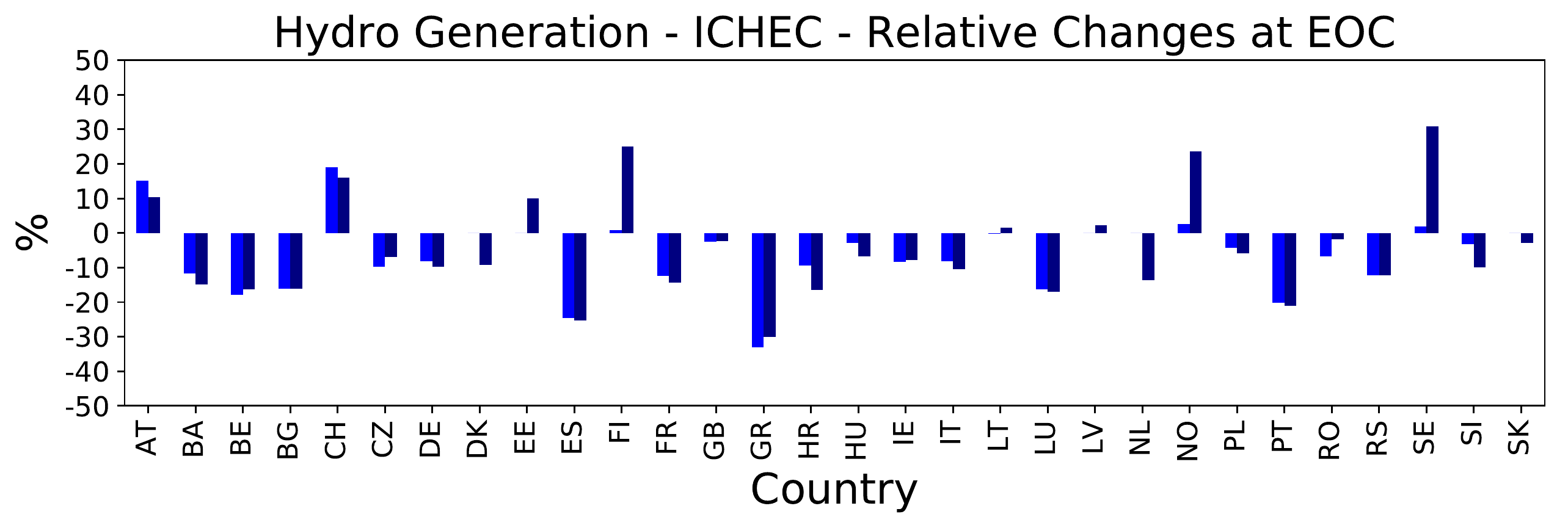}
\includegraphics[width=.9\textwidth]{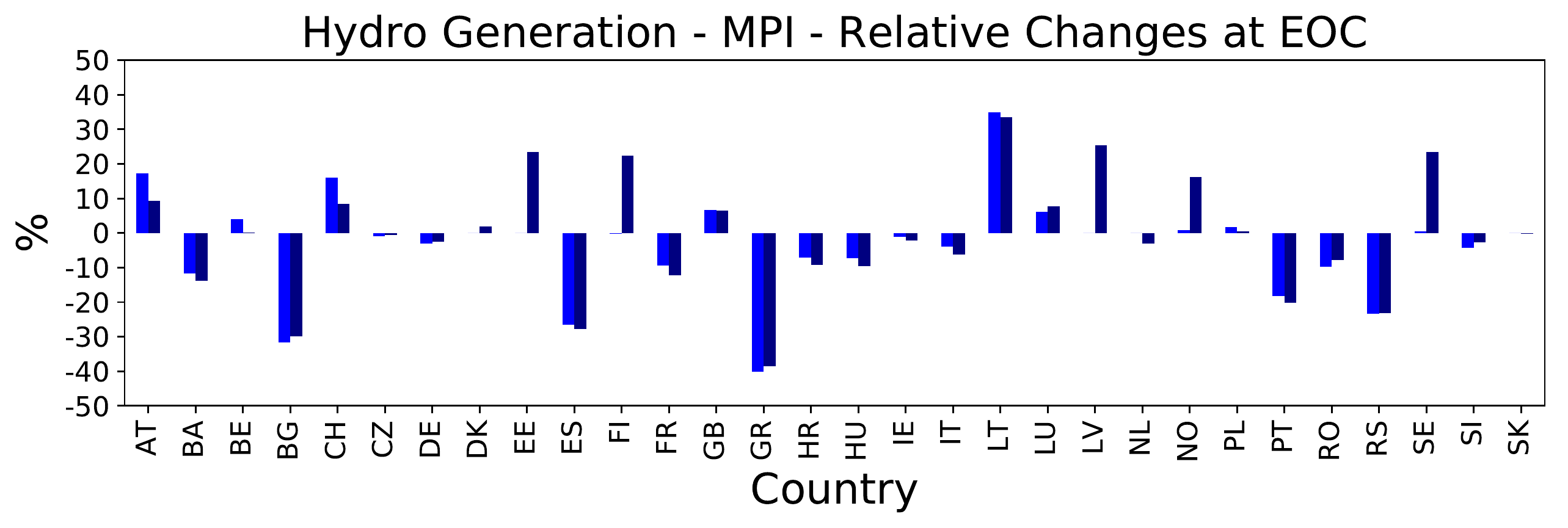}
\caption{\label{fig:hydro_generation_relative} Continuation of the last figure. Shown are the relative changes in reservoir and run-off-river generation at the EOC period compared to the corresponding HIS periods.}
\end{center}
\end{figure}

Fig. \ref{fig:hydro_generation_absolute} and \ref{fig:hydro_generation_relative} show the actual average hydro power generation per country and its relative changes. Hydro power generation is dependent on the weather-driven inflow shown in the resource analyses in Sec \ref{sec:res}. In addition, hydro power provides large storage and dispatch capabilities. Both, reservoir and run-off-river plants show a quite similar behaviour since they rely on the same renewable resource, i.e. total water runoff. Hence, the trend in generation for both carrier types evolves qualitatively in the same direction.


For the different models, hydro generation strongly fluctuates in the transition from the HIS period to the EOC period. While in northern Europe small increases are observable, dispatch in southern European countries decreases by a large degree. It seems likely that observed decreases in hydro generation are a driver for PV installations in the southern and south-eastern countries until the end of century.


\subsection{Total System Costs}
\begin{figure}
\begin{center}
\includegraphics[width=.4\textwidth]{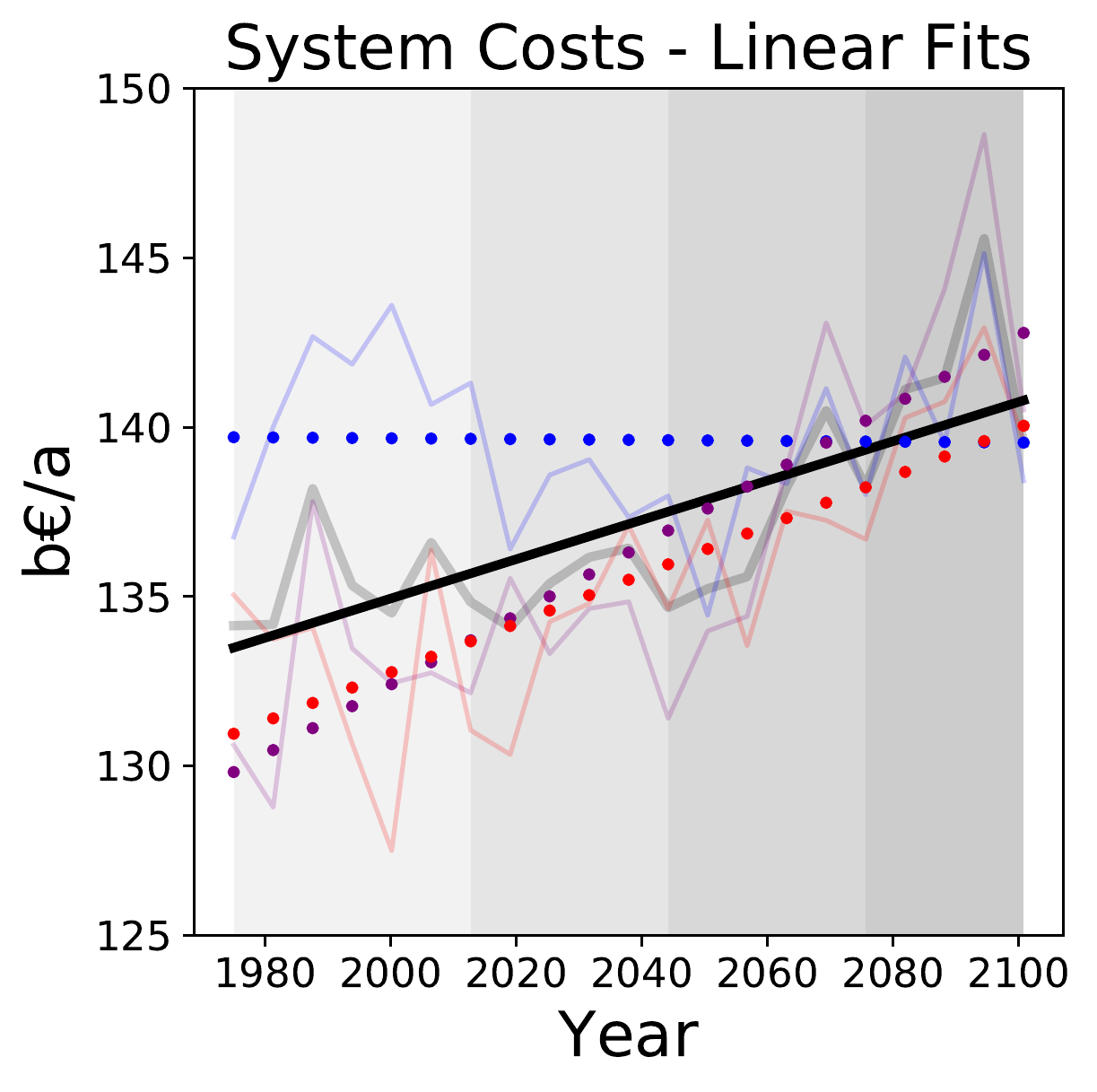}\\
\includegraphics[width=.4\textwidth]{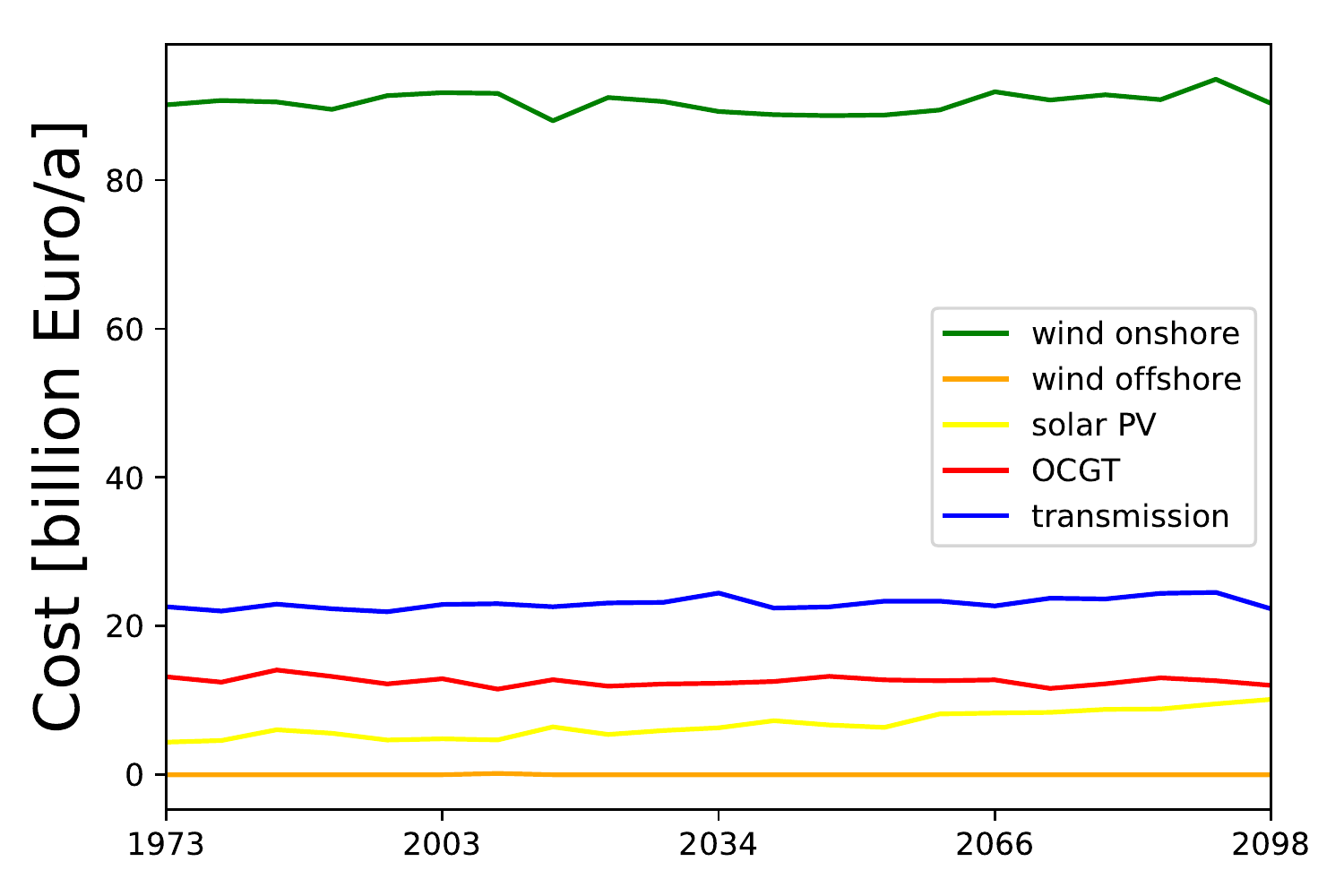}
\includegraphics[width=.4\textwidth]{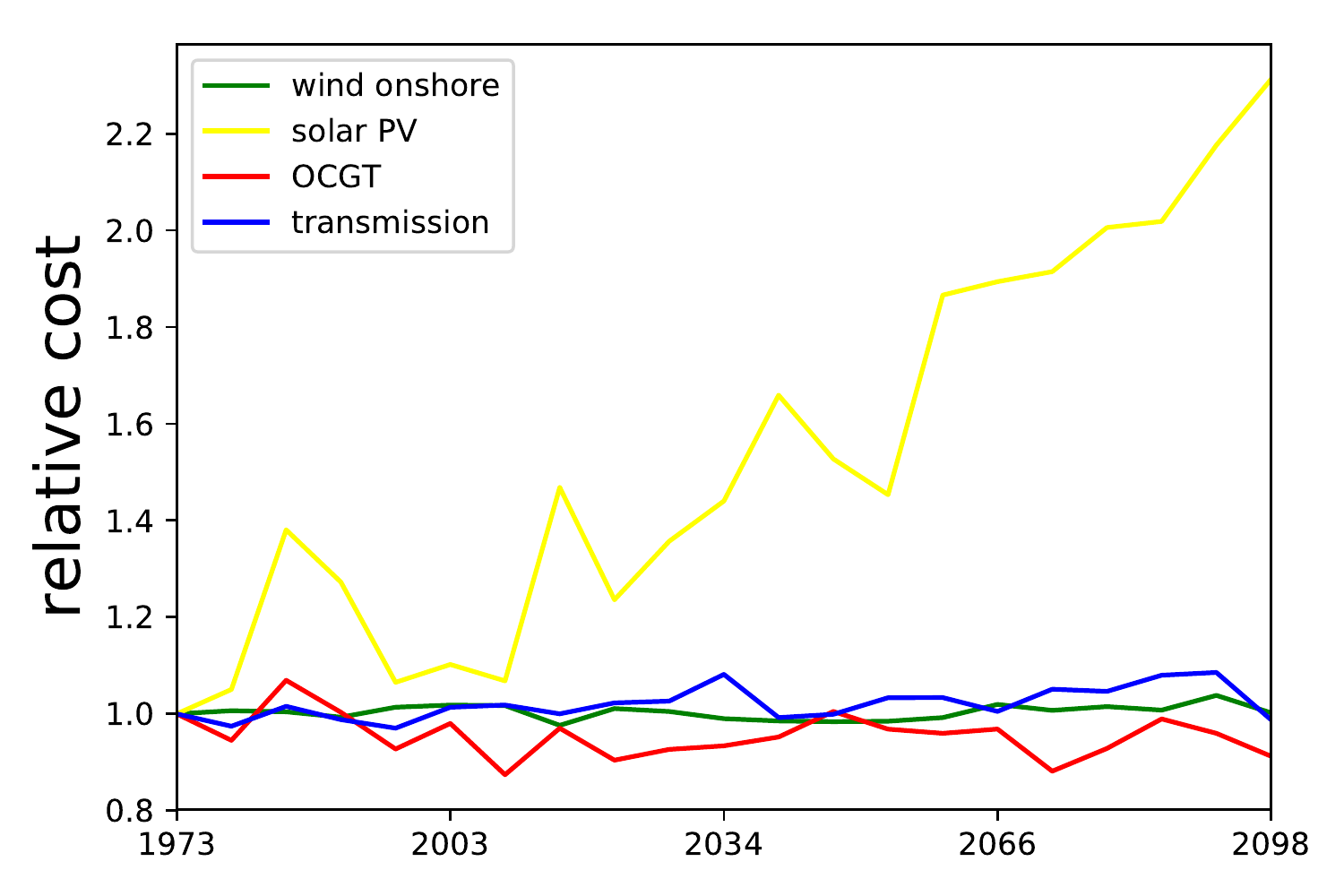}
\caption{\label{fig:system_costs} Total system costs in the cost-optimal case for the ensemble mean as well as the single ensemble members.}
\end{center}
\end{figure}

Fig. \ref{fig:system_costs} shows the cost-development for the different models over time.
The ensemble mean of the cost-optimal solution predicts total system costs of about 135.5 billion Euro at the HIS period and 141.2 billion Euro at the EOC period, which amounts to an increase in total system costs of about 4\%. Using a linear fit, this increase is represented by a corresponding slope of about 360 million Euro p.a., which represents the cost-impact of climate change on the European power system in the given renewable setting under a greenfield investment model. This increase is caused by decreasing resource quality and increasing correlation lengths for wind, which in turn increases the need for additional balancing measures.
Almost the entire increase in costs can be attributed to the increase of expenditures for solar PV, where costs increase by 5.8 billion Euro p.a..
Cost for onshore wind and transmission remain virtually unchanged, while cost of OCGT is even reduced by 1.2 billion Euro p.a..
The evaluation of the single models indicate rising system costs until the end of century as a general feature, where total cost increases. While the MPI and ICHEC model predict a large cost increase with a slope between 450 and 650 million p.a. starting at low total base costs of about 130 billion Euro, the CNRM model in opposite forecasts only small cost increases, but starts at already high total base system costs of about 140 billion Euro (Table \ref{table:cost_fits}). Finally, at the EOC period the fits converge at a value of about 140 billion Euro.

\begin{table}[!ht]
\centering
\begin{tabular}{||c|c|c|c|c||}
\hline
\textbf{Model} & \textbf{Slope} & \textbf{Slope Error} & \textbf{Axis Intercept} & \textbf{Axis Intercept Error}\\
& [bEuro/a] & [bEuro/a] & [bEuro] & [bEuro]\\
\hline
Ensemble & 0.36 & 0.08 & 133.49 & 0.88\\
CNRM & 0.01 & 0.01 & 139.71 & 1.12\\
ICHEC & 0.65 & 0.11 & 129.82 & 1.32\\
MPI & 0.45 & 0.09 & 130.95 & 1.08\\
\hline
\end{tabular}
\caption{Fit parameters obtained from the fits for total system costs in the cost-optimal case. Fits are shown in Fig. \ref{fig:system_costs}.}
\label{table:cost_fits}
\end{table}

\section{Discussion}
\label{sec:discussion}
The results of this paper might be effected by a number of simplifications and remaining uncertainties:

\begin{itemize}
\item 1) the optimisation result is dominated by generation from onshore wind and contains no or only minor shares of offshore wind generation. However, costs for offshore wind installations have fallen in recent years much faster than predicted and hence might not be adequately reflected by the cost assumptions. In general, cost assumptions from Table \ref{tab:costsassumptions} might change significantly in a few years due to advancements in technology, especially considering recent developments in PV and storage technologies.

\item 2) the network setup consists of 30 nodes only. The optimum in a finer network consisting of much more nodes may show other structures of installed capacities and transmission lines. 

\item 3) load data consists of historical values accessed from ENTSO-E. However, it is expected that climate change also affects the electricity demand, so climate change affected demand data could be included as well.

\item 4) climate projections suffer from inherent modelling uncertainties, especially considering the irradiation models, whose energy balance is strongly coupled to atmospheric dynamics, evaporation and clouding processes and hence the wind speed and water runoff variable. The optimised power system in turn is strongly coupled to the resource quality and reacts sensibly to those climate modelling issues.
\end{itemize}

The sensitivity towards cost assumptions is discussed by Schlachtberger et al. \cite{schlachtberger2018cost} using the same methodology as in this paper but different weather data.
They find that changes in investment cost of wind/PV technologies have significant effect on the cost-optimal mix and distribution of renewable generation.
However, since it is likely that the ratio of investment cost over capacity factor plays a decisive role, this can also be interpreted in terms of this paper: Because climate change affects the ratio of generation costs between renewable generation technologies, it directly effects the cost-optimal mix and distribution. 
In addition, they study the influence of different consecutive weather years as input data in the model and find that different years results in different cost-optimal generation mixes, which is likely caused by year-to-year fluctuations in renewable capacity factors, but overall system costs are only weakly affected.

Hoersch and Brown \cite{horsch2017role} study the effect of spatial resolution in the methodological setup used in this paper and find that the effect of spatial resolution is strongest emphasized with respect to inter-connecting transmission and also has observable effect on overall system costs, if the transmission system is strongly reinforced. This can be understood considering that increasing the spatial resolution gives more variety placing renewable generation sites and thus more options for cost-optimal placement.

Concerning data quality of the EURO-CORDEX project,
Tobin et al. \cite{tobin2016climate} compare its near-surface wind speed data against observations from ISDLite \cite{smith2011integrated} and QuikSCAT \cite{ruti2008comparison} and find reasonable wind speed variability from the EURO-CORDEX model with some detected problems over land.
Jerez et al. \cite{jerez2015impact} evaluate radiation from EURO-CORDEX ensemble members and attest a good agreement with spatial and temporal variability of radiation patterns. 
In addition, it should be noted that climate models tend to underestimate intermultidecadal variations in shortwave radiation \cite{allen2013evaluation}. 

Other studies have tried to assess the impact of climate change on a renewable power system as well.
Wohland et al. \cite{wohlandmore} focus, using EURO-CORDEX ensemble members, on wind alone and find the quality of wind power to decrease, i.e., spatial correlation to increase and low wind situations to worsen. Both effects likely contribute to the increasing share of PV generation in the cost-optimal case, as observed by our study.
Weber et al. \cite{weber2017impact} use the data and a simplified model of a European power system, where investment and dispatch are not cost-optimal, but instead derived using heuristic approaches (minimising needs for certain quantities such as backup energy). They find an increase of the likelihood for long periods of low wind generation and, in addition,
an increase in the seasonal wind variability
Kozarcanin et al. \cite{kozarcanin2018climate} define different key metrics such as transmission benefits in a simplified setting to quantify the impact on climate change on a power system and conclude.
They find the benefit of transmission to shrink due to climate change and the need for backup energy to increase.
This is again likely associated with decreased wind quality, because wind and transmission are natural complementarities due to low spatial long-distance correlation of wind feed-in.
All results of these studies are in line with our major finding concerning the cost-optimal generation mix: the importance of wind will shrink, the importance of solar PV will increase.

\section{Conclusions}
\label{sec:conclusion}
As discussed throughout this paper, future climate-driven changes in renewable resources will likely have an impact on renewable power systems around the world.
We have studied energy-related variables of climate-affected weather datasets (EURO-CORDEX) and their implications for the design of a future cost-optimal renewable European power system.
The analysis allows for investment advice to be made despite an uncertain climate future. From the presented results, we draw the following conclusions:

\begin{itemize}
\item climate change has an effect on the optimal power system structure. However, it should always be considered that climate change models do exhibit uncertainties, which can also clearly be seen by comparison between different models.
The decision, whether climate-affected or historical weather data should be used to model renewable power systems of the (far and not so far) future is delicate, as both effects must be assigned meaningful weights.

\item the three ensemble members, employing different general circulation models but one concurring regional climate model, differ in some aspects and show strong similarities in others: Most importantly they differ in the projection of wind speeds and resulting wind capacity factors, while they show very similar results for incoming irradiation and corresponding PV capacity factors as well as qualitatively opposite behaviour of water runoff and the following hydro generation.

\item wind, PV and hydro resources are all affected by global warming, but consequences for the cost-optimal renewable European power system differ: For wind and PV, a clear north-south divide can be identified, where onshore wind is favoured in the north and remains stable until the end of century and PV is favoured in the south gaining increased shares at the end of century. Hydro generation in turn shows small increases in the north, but suffers from losses in southern Europe and is apparently replaced by PV. The cost-optimal renewable European power system requires an extension of today's value of inter-connecting transmission line capacities (30 TWkm) by a factor of around 9-10, which grows by a few percent until the end of century.

\item total system costs increase until the end of century; the different climate models contradict each other in this point, but converge at the end of century. 
\end{itemize}

The core observion of this study is that climate change affects the optimal infrastructure of a European renewable power system in an observeable way: The role of PV grows and the need for balancing infrastructure alongside. This was observed as a result of all models studied and it is therefore assumed to be the most stable and probable impact of climate change. 
The major conclusion is therefore that effects due to climate change should be considered or, if this puts to much stress on computational cost, at least be discussed in long-term renewable power system studies. 
The same holds true for siting of renewables, where not only changes in resource quality due to climate change should be taken into account but also the effects of climate change on the overall power system (e.g., the cost-optimal distribution of renewables).

\section*{Acknowledgment}
This research was conducted as part of the CoNDyNet project, which is supported by the German Federal Ministry of Education and Research under grant no. 03SF0472C. 
Alexander Kies is financially supported by Stiftung Polytechnische Gesellschaft Frankfurt am Main.
Martin Greiner is partially funded by the RE-INVEST project (Renewable Energy Investment Strategies -- A two-dimensional interconnectivity approach), which is supported by Innovation Fund Denmark (6154-00022B).
Tom Brown acknowledges funding from the Helmholtz Association under grant no. VH-NG-1352.
The responsibility for the contents lies solely with the authors.
Parts of this work were presented at the Wind Integration Workshop 2017 \cite{schlott2017impact}.

\bibliographystyle{plain}
\bibliography{./paper.bib}

\end{document}